\documentclass[aps,prd,superscriptaddress,nofootinbib,11pt]{revtex4}
\usepackage[english]{babel}
\usepackage[utf8]{inputenc}
\usepackage{graphicx}   % need for figures
\usepackage{slashed}
\usepackage{epstopdf}
\usepackage{verbatim}   % useful for program listings
\usepackage{color}      % use if color is used in text
\usepackage{subfigure}  % use for side-by-side figures
\usepackage{multirow}
\usepackage{hyperref}   % use for hypertext links, including those to external documents and URLs
\usepackage{float}
\usepackage{epsfig,rotating}
\usepackage{amsmath,amssymb}
\usepackage{dsfont}
\restylefloat{table}
\numberwithin{equation}{section}

\newcommand{\vx}{\vec{x}}

\newcommand{\vk}{\vec{k}}

\newcommand{\be}{\begin{equation}}
\newcommand{\ee}{\end{equation}}
\newcommand{\bea}{\begin{eqnarray}}
\newcommand{\eea}{\end{eqnarray}}

\newcommand{\ket}[1]{|#1\rangle}
\newcommand{\bra}[1]{\langle#1|}

\newcommand{\ta}{\widetilde{A}}
\newcommand{\tb}{\widetilde{B}}
\newcommand{\tf}{\widetilde{f}}

\newcommand{\X}{\varphi}

\begin{document}

\title{Is the finite temperature effective potential, effective for dynamics?.}

\author{Nathan Herring}
\email{nherring@hillsdale.edu} \affiliation{Department of Physics, Hillsdale College, Hillsdale, MI 49242}
\author{Shuyang Cao}
\email{shuyang.cao@pitt.edu} \affiliation{Department of Physics, University of Pittsburgh, Pittsburgh, PA 15260}
 \author{Daniel Boyanovsky}
\email{boyan@pitt.edu} \affiliation{Department of Physics, University of Pittsburgh, Pittsburgh, PA 15260}

\date{\today}

\begin{abstract}
  We study the applicability of the finite temperature effective potential in the equation of motion of a
  homogeneous ``misaligned'' scalar condensate $\varphi$, and find important caveats that severely restrict its domain of validity: \textbf{i:)} the \emph{assumption} of local thermodynamic equilibrium (LTE) is in general not warranted, \textbf{ii:)} we show a direct relation between the effective potential and the thermodynamic entropy density $\mathcal{S}= - \partial V_{eff}(T,\varphi)/\partial T$, which entails that for a dynamical $\varphi(t)$ the entropy becomes a non-monotonic function of time, \textbf{iii:)}  parametric instabilities in both cases with and without spontaneous symmetry breaking lead to profuse particle production with
  non-thermal distribution functions, \textbf{iv:)} in the case of spontaneous symmetry breaking   spinodal instabilities yield a complex effective potential, internal energy and \emph{entropy}, an untenable situation in thermodynamics. All these caveats associated with using the effective potential in the \emph{equation of motion} of the condensate, cannot be overcome by finite temperature equilibrium resummation schemes.
We argue that the dynamics of the condensate leads to decoupling and freeze-out from (LTE), and  propose a \emph{closed} quantum system approach based on unitary time evolution. It  yields  the correct equations of motion without the caveats of the effective potential, and provides a fully renormalized and thermodynamically consistent framework to study the dynamics of the ``misaligned'' condensate, with real and conserved energy and entropy amenable to numerical study. The evolution of the condensate leads to profuse   \emph{stimulated particle production} with non-thermal distribution functions. Possible emergent asymptotic non-thermal states and eventual re-thermalization are conjectured.
\end{abstract}


\maketitle

\section{Introduction}
The finite temperature effective potential is a very powerful diagnostic tool to study the phase structure of quantum field theories including thermal and quantum corrections. It is the finite temperature extension of the zero temperature effective potential originally proposed by the pioneering work of refs.\cite{jona,goldstone,colewein,veffsy} to study how radiative corrections modify the symmetry breaking properties of the vacuum. Functional methods provide a systematic formulation of the zero temperature effective potential as the generating functional of single particle irreducible Green's functions at \emph{zero four momentum}\cite{jackiw1,ilio,colepoli,colemanbook}.

The extension of the effective potential to equilibrium finite temperature was pioneered by refs.\cite{jackiw,weinberg}. It describes the free energy landscape as a function of the  spatially homogeneous and time independent order parameter,  the expectation value of a scalar field $\varphi$, thereby characterizing the different phases of a theory. As such, the finite temperature effective potential plays a fundamental role in cosmology, as it may describe possible cosmological phase transitions\cite{kolb,branrmp,guth,linde,albrecht}.

\textbf{Motivation and Objectives:} The finite temperature effective potential was originally introduced and developed with the aim of describing \emph{equilibrium} aspects of spontaneous symmetry breaking including quantum and thermal effects in terms of a free energy as a function of the homogeneous and static order parameter. However, it is often used in the equations of motion of such an order parameter to describe the \emph{dynamics} of, for example, ``misaligned'' condensates. A recent study\cite{veff} of the zero temperature effective potential, extending the Hamiltonian framework introduced in refs.\cite{veffsy,weinbergwu}, revealed several important caveats that invalidate its applicability in the equation of motion of the order parameter, namely the  condensate or mean field. In the case when the tree level potential does not feature broken symmetry minima, oscillations of the condensate around its  minimum lead to instabilities associated with parametric amplification resulting in  the exponential growth of the fluctuations around the mean field with profuse particle production, a physical mechanism similar to that of reheating in cosmology\cite{rehe1,rehe2,rehe3,branreh,rehe4,rehe5,rehe6}. In the case when the tree level potential admits broken symmetry minima, a different instability emerges when the excursion of the mean field probes a region where the potential features negative second derivatives. This is the spinodal (or tachyonic) instability and again leads to exponential growth of fluctuations around the mean field. In this case the growth of fluctuations is associated with the formation and growth of correlated domains\cite{weinbergwu,calzetta,boyaspino}. In statistical physics this is the hallmark of the process of spinodal decomposition and phase ordering dynamics in phase transitions\cite{langer1,langer2,allen,gunton}.

Both types of instabilites lead to the unambiguous conclusion that the zero temperature effective potential, which by definition and construction is a \emph{static} function of the mean field, is inadequate to describe the dynamics of the mean field\cite{veff}.

Motivated by the ubiquity and importance of the effective potential in cosmology and in general to study the phases of a quantum field theory, our objectives in this article are: \textbf{i:)}  to critically examine the validity of the finite temperature  effective potential in the \emph{equation of motion} of a homogeneous ``misaligned'' condensates, \textbf{ii:)} if it is found to   be unreliable,  to provide an alternative and  consistent formulation of the dynamics of the condensate. While ultimately our aim is to study these aspects within the context of an expanding cosmology, in this article we restrict our focus to the case of Minkowski space-time as a first step. Undoubtedly, a critical assessment of the validity of the effective potential in the equation of motion of condensates must start with this simpler case from which much can be learned and whose study will pave the way towards a firmer understanding in cosmology.

\vspace{1mm}

\textbf{Brief summary of results:} We extend and complement the Hamiltonian formulation of the finite temperature effective potential introduced in ref.\cite{weinberg},  yielding   a clear relation to the zero temperature case studied in ref.\cite{veff}. We  obtain an   exact result: the finite temperture effective potential is the Helmholtz free energy density for the \emph{fluctuations} around the expectation value of the scalar field $\varphi$ (order parameter).   This relation has an important thermodynamic consequence: $\mathcal{S} = -\partial \, V_{eff}[T,\varphi]/\partial T$,  where $\mathcal{S}$ is the thermodynamic entropy density. Therefore, the applicability of the   effective potential  in a dynamical equation of motion for a ``misaligned'' condensate $\varphi$ is restricted by fundamental thermodynamic properties of the entropy. In the case of unbroken symmetry we find a non-monotonic time dependence of the entropy, and in the case of broken symmetry the entropy becomes \emph{complex} as a consequence of spinodal instabilities. Both cases are untenable in local thermodynamic equilibrium. Implementing a Chapman-Enskog expansion of the Boltzmann equation, we show that the assumption of local thermodynamic equilibrium (LTE) is in general unwarranted as it requires a fine tuning of couplings to the heat bath. Furthermore, we argue that parametric and spinodal instabilities invalidate the use of an effective potential, which by design and construction is a \emph{static} equilibrium function of $\varphi$, in the equation of motion for the condensate. It is argued that the dynamical evolution of the condensate leads to a ``freeze out'' of the density matrix and decoupling from (LTE), and propose a closed quantum system approach to the dynamics. We introduce a method based on unitary time evolution to obtain directly the correct equations of motion for the condensate, which features   conservation of   energy   and entropy,  these are always real and without the caveats associated with the effective potential. Parametric and spinodal instabilities lead to an energy transfer between the condensate and the fluctuations resulting in \emph{stimulated} particle production with non-thermal distribution functions. A fully renormalizable and thermodynamically consistent framework to study the dynamics, amenable to   numerical study is provided. Possible asymptotic states and re-thermalization are conjectured.

The article is organized as follows: in section (\ref{sec:static}), we briefly review the Hamiltonian approach to the zero temperature effective potential before extending the formulation from ref.\cite{weinberg} to finite temperature. In this section we relate the static effective potential non-perturbatively to the Helmholtz free energy and thermodynamic entropy of the fluctuations and obtain the well known result for the one loop effective potential. In section (\ref{sec:dynamics}) we analyze the reliability of the static effective potential in the equation of motion for the condensate under the assumption of (LTE). In this section we show that (LTE) is in general not warranted and discuss severe caveats in the use of the effective potential in the equations of motion arising from parametric amplification and spinodal instabilities. In section (\ref{sec:closed})  it is argued that the dynamics leads to a ``freeze out'' from (LTE) and decopuling from thermal bath and introduce a \emph{closed} quantum system approach based on unitary time evolution  to obtain the correct equations of motion. These are shown to  conserve  energy density and entropy,  which are manifestly real without the caveats of the effective potential. In this section it is shown that parametric and spinodal instabilities are efficient mechanisms of energy transfer between the condensate and the fluctuations leading to profuse \emph{stimulated} particle production with non-thermal distributions.We provide a fully renormalized and thermodynamically consistent framework to study the dynamics of the condensate amenable to numerical study. Section (\ref{sec:discussion}) conjectures on the emergence of possible asymptotic states and re-thermalization. In section (\ref{sec:conclusions})  we present our conclusions and suggest further avenues of study. An appendix summarizes the non-equilibrium correlation functions needed to obtain the equations of motion.

\section{The static  effective potential: zero vs finite temperature}\label{sec:static}
\subsection{Zero temperature}\label{subsec:zero}

Before we consider the finite temperature effective potential, we briefly summarize the main concepts behind the Hamiltonian approach of refs.\cite{veffsy,weinbergwu,veff} with the objective of comparing with the finite temperature case discussed below.

Let us consider a scalar theory described by the Hamiltonian
 \be \hat{H} = \int d^3x \Bigg\{\frac{\hat{\pi}^2}{2}+\frac{(\nabla \hat{\phi})^2}{2}+V(\hat{\phi})\Bigg\}\;. \label{Hamiltonian1}\ee where $\hat{\pi}$ is the canonical momentum conjugate to the scalar field $\hat{\phi}$. The Hamiltonian interpretation of the effective potential\cite{veffsy,weinbergwu,veff} defines the effective potential as the expectation value of $\hat{H}$ in a normalized \emph{coherent state}  $\ket{\Phi}$ in which the field acquires a \emph{space-time independent expectation value}, namely a \emph{mean field} $\varphi$,
    \be \varphi = \bra{\Phi}\hat{\phi}(\vx,t)\ket{\Phi}~~;~~\bra{\Phi}\hat{\pi}(\vx,t)\ket{\Phi} =0\,,\label{statexvals}\ee as
    \be V_{eff}(\varphi) \equiv \frac{1}{\mathcal{V}}\,\bra{\Phi}\hat{H}(\vx,t)\ket{\Phi}\,,\label{veffstatic} \ee where $\mathcal{V}$ is the spatial volume. Shifting the field operator $\hat{\phi}$ by its coherent state expectation value $\varphi$,
     \be \hat{\phi}(\vx,t) = \varphi+ \hat{\delta}(\vx,t)~~;~~ \hat{\pi}(\vx,t) \equiv \hat{\pi}_\delta(\vx,t)\,, \label{fieldsplit}\ee the constraints (\ref{statexvals}) imply
    \be \bra{\Phi}\hat{\delta}(\vx,t)\ket{\Phi}=0 ~~;~~\bra{\Phi}\hat{\pi}_\delta(\vx,t)\ket{\Phi} =0\,,\label{statexvalsdel}\ee leading to
    \be V_{eff}  = \,V(\varphi)+ \frac{1}{\mathcal{V}}\, \int d^3x \,\bra{\Phi}\Big\{\frac{\hat{\pi}^2_\delta}{2}+\frac{(\nabla \hat{\delta})^2}{2}+\frac{1}{2}\,V''(\varphi)\,\hat{\delta}^2 + \cdots \Big\}\ket{\Phi}\;,  \label{Hamiltonian2} \ee where the expectation value of the linear terms in $\hat{\delta}$ and $\hat{\pi}_\delta$ vanish by the constraints (\ref{statexvals}), and the dots in eqn. (\ref{Hamiltonian2}) stand for higher powers of $\delta$ leading to higher loop corrections.

In the Hamiltonian formulation  quantization proceeds by expanding the fluctuation field $\delta(\vx,t)$ in the basis of solutions of the   Heisenberg field equations for a free field with mass squared $V^{''}[\varphi]$, namely
\be \ddot{\delta}(\vx,t)-\nabla^2\,\delta(\vx,t) + V^{''}[\varphi]\,\delta(\vx,t) =0\,,\label{heis}\ee  and the field $\delta(\vx,t)$ is expanded in mode functions,
\bea \delta(\vx,t) & = & \frac{\sqrt{\hbar}}{\sqrt{\mathcal{V}}}\sum_{\vk}\Big[ a_{\vk}\, g_k(t)\,e^{i\vk\cdot\vx} + a^\dagger_{\vk}\,g^*_k(t)\,e^{-i\vk\cdot\vx}\Big]\,,\label{delex}\\
\pi_{\delta}(\vx,t) & = & \frac{\sqrt{\hbar}}{\sqrt{\mathcal{V}}}\sum_{\vk}\Big[ a_{\vk}\, \dot{g}_k(t)\,e^{i\vk\cdot\vx} + a^\dagger_{\vk}\,\dot{g}^*_k(t)\,e^{-i\vk\cdot\vx}\Big]\,,\label{pidelex}
\eea

where  the mode functions $g_k(t)$ are solutions of the equations\cite{veff}
\be \ddot{g}_k(t) + \omega^2_k(t)\,g_k(t) =0 ~~;~~ \omega^2_k = k^2+V^{''}[\varphi]\,,\label{modeg}\ee
 with the Wronskian condition
 \be \dot{g}_k(t)\,g^*_k(t) - \dot{g}^*_k(t)\,g_k(t) = -i \,,\label{wron}\ee  and the annihilation and creation operators are time independent and obey canonical commutation relations. For a \emph{space-time constant} $\varphi$, the mode functions are given by
 \be g_k(t) = \frac{e^{-i\omega_k t}}{\sqrt{2\omega_k}} ~~;~~ \omega_k = \sqrt{k^2+V^{''}[\varphi]}\,,\label{gkficons}\ee yielding the mode expansion
\bea \delta(\vx,t) & = & \frac{\sqrt{\hbar}}{\sqrt{\mathcal{V}}}\sum_{\vk}\frac{1}{\sqrt{2\omega_k}}\Big[ a_{\vk}\, e^{-i\omega_k t}\,e^{i\vk\cdot\vx} + a^\dagger_{\vk}\,e^{i\omega_k t}\,e^{-i\vk\cdot\vx}\Big]\,,\label{quandelta}\\
\pi_{\delta}(\vx,t) & = & -i\frac{\sqrt{\hbar}}{\sqrt{\mathcal{V}}}\sum_{\vk}\sqrt{\frac{\omega_k}{2}}\Big[ a_{\vk}\, e^{-i\omega_k t}\,e^{i\vk\cdot\vx} - a^\dagger_{\vk}\,e^{i\omega_k t}\,e^{-i\vk\cdot\vx}\Big]\,,\label{quanpidelta}
\eea and
    the quadratic Hamiltonian inside the brackets in eqn. (\ref{Hamiltonian2})   becomes
    \be H = \sum_{\vk} \hbar \omega_k(\varphi) \Big[a^{\dagger}_{\vk}\,a_{\vk}+\frac{1}{2}\Big]\,. \label{hamqua}\ee

    The constraints (\ref{statexvalsdel}) are implemented by requesting that   the coherente state $\ket{\Phi}$ be an eigenstate of the Fock occupation number $n_{\vk} = a^\dagger_{\vk}\,a_{\vk}$, however the lowest expectation value of the quadratic Hamiltonian is obtained for the \emph{vacuum state} for the  fluctuations $\hat{\delta}$, namely\cite{veff}
    \be a_{\vk}\ket{\Phi} =0 ~,~ \forall \vk \,, \label{anni}\ee leading to the constraint (\ref{statexvalsdel}).

     Taking the infinite volume limit with $\sum_{\vk} \rightarrow \mathcal{V}\,\int d^3k/(2\pi)^3$ and using (\ref{anni}), we find that  the effective potential (\ref{veffstatic}) is given by
    \be V_{eff}(\varphi) = V(\varphi)+ \frac{\hbar}{2} \int \frac{d^3k}{(2\pi)^3}\,\omega_k(\varphi) + \mathcal{O}(\hbar^2) +\cdots \,.\label{Veff1lupst}\ee The $\hbar$ in (\ref{Veff1lupst}) originates in the $\sqrt{\hbar}$ in the usual field quantization (\ref{quandelta},\ref{quanpidelta}) and implies that the expression (\ref{Veff1lupst}) is the zero temperature \emph{one loop effective potential}\cite{veff,weinbergwu}. If $\ket{\Phi}$ is an excited eigenstate with $n_k\neq 0$, the integrand in the second term features an extra contribution $n_k\,\omega_k(\varphi)$ thereby raising the energy.

    In order to compare the above results to the finite temperature case, we introduce the pure state density matrix
    \be \rho \equiv \ket{\Phi}\bra{\Phi} \,,\label{denmtxzero}\ee from which it follows that
    \be \varphi = \mathrm{Tr}\hat{\phi}(\vx) \,\rho ~~;~~ V_{eff}(\varphi) = \frac{1}{\mathcal{V}}\, \mathrm{Tr}H \,\rho\,,\label{exvalszero}\ee and the constraints (\ref{statexvalsdel}) become
    \be  \mathrm{Tr}\hat{\delta}(\vx) \,\rho =0 ~~;~~  \mathrm{Tr}\hat{\pi}_{\delta}(\vx) \,\rho =0 \,.\label{consdmtx}\ee

Before moving on to the finite temperature case, it must be emphasized the there are two  main \emph{assumptions} leading up to the zero temperature result (\ref{Veff1lupst}): \textbf{i:)} that the mean field $\varphi$ is \emph{time independent}, yielding  the mode functions given by eqn. (\ref{gkficons}),   \textbf{ii:)} that $V^{''}(\varphi)>0$, condition that yields \emph{real} frequencies $\omega_k(\varphi)$.

When $\varphi$ evolves in time, as in the dynamical case,  the mode functions $g_k(t)$ are solutions of the mode equations (\ref{modeg}) with $V^{''}(\varphi(t))$ that now depends on time through the time dependence of $\varphi(t)$, and gives rise to parametric instabilities, and if $V^{''}(\varphi)$ is \emph{negative} for some values of $\varphi$, there are instabilities for $k^2<|V^{''}(\varphi)|$ since for these wavevectors the frequencies $\omega_k$ become purely imaginary. Both instabilities will be addressed in section (\ref{sec:dynamics}) within the context of the applicability of the finite temperature effective potential in the dynamical evolution of $\varphi(t)$ and have been discussed in greater depth   in ref.\cite{veff} to which we refer the reader for a more detailed treatment.

\subsection{Finite temperature}\label{subsec:finiteT}

The discussion above highlights the interpretation of the zero temperature effective potential as the expectation value of the Hamiltonian in a particular coherent state, defined to be the vacuum for the fluctuations around the mean field $\varphi$. This formulation does not have a straightforward extrapolation to finite temperature, because the equilibrium density matrix corresponds to  a mixed state that describes an ensemble, not a pure state  as in (\ref{denmtxzero}). The constraints (\ref{statexvalsdel}), which can be implemented straightforwardly in the case of a pure state,  must now be imposed in terms of Lagrange multipliers added to the Hamiltonian in the thermal density matrix. This is achieved by following the formulation of the finite temperature effective potential advocated in the seminal articles \cite{jackiw,weinberg}. In particular
by implementing the Hamiltonian formulation of ref.\cite{weinberg}\footnote{See the appendix in this reference.}, wherein the effective potential is obtained from the Legendre transform
of the equilibrium free energy under the constraint that the expectation value of the field is a space-time constant.

To discuss the main arguments in a clear manner, we focus on the simple case of a scalar field with a Lagrangian density
\be \mathcal{L}[\phi] = \frac{1}{2} \partial_{\mu}\phi\partial^\mu\phi - V[\phi] \,,\label{lagran}\ee yielding the Hamiltonian (\ref{Hamiltonian1}).  Let us introduce
\be H_J[\phi] = H[\phi]+\int d^3x J(\vx)\,\phi(\vx)\,,\label{hamj}\ee where $J(\vx)$ is an external classical source. The canonical partition function is given by
\be \mathcal{Z}[T;J] \equiv  e^{-\beta F[T;j]} = \mathrm{Tr}\, e^{-\beta H_J[\phi]} \,, \label{ZJ} \ee where $F[T;J]$ is the Helmholtz free energy and $\beta = 1/T$. The \emph{equilibrium} expectation value of the field $\phi(\vx)$ in presence of the source is defined as
\be \varphi(\vx) \equiv \frac{\mathrm{Tr} \,\phi(\vx)\, e^{-\beta H_J[\phi]}}{\mathrm{Tr}\, e^{-\beta H_J[\phi]}} \,,\label{expfi} \ee which is obtained as a variational derivative with respect to the c-number source, namely
\be \varphi(\vx) = -\frac{1}{\beta} \frac{\delta}{\delta J(\vx)} \ln\big[\mathcal{Z}\big] =  \frac{\delta}{\delta J(\vx)} F[T;J] \,, \label{expfi2}\ee  the source $J(\vx)$ can be interpreted as a Lagrange multiplier for the constraint  $\varphi(\vx) = \langle \phi(\vx) \rangle$ where the expectation value is obtained with the partition function $\mathcal{Z}[T;J]$.  The relations (\ref{expfi},\ref{expfi2})  are inverted to yield
\be J(\vx) = J[\varphi(\vx)]\,,\label{jfi} \ee from which the Legendre transform
\be \Omega[T;\varphi] = F[T;J[\varphi]]- \int d^3 x J[\varphi(\vx)]\,\varphi(\vx) \,,\label{Gibbs}\ee yields the generalized Gibbs free energy as a function(al) of
temperature and the expectation value $\varphi(\vx)$. Using the definition (\ref{expfi}) it is straightforward to find that
\be \frac{\delta}{\delta \varphi(\vx)} \Omega[T;\varphi] = - J[\varphi(\vx)] \,. \label{derifi}\ee  From now on we will consider a spatially constant expectation value $\varphi$, which implies a translationally invariant partition function,  and introduce
\be j \equiv \frac{1}{\mathcal{V}} \, {\int d^3 x J(\vx) } ~~;~~ F[T;J] \equiv \mathcal{V} \,\mathcal{F}[T;j]\,,\label{defs} \ee with $\mathcal{V}$ the spatial volume, and following references\cite{weinberg,jackiw}  \emph{define}  the finite temperature effective potential as
\be \Omega[T;\varphi] \equiv \mathcal{V}\,V_{eff}(T;\varphi)\,\label{Veff} \ee From equations (\ref{Gibbs},\ref{Veff})  it follows that
\be V_{eff}[T;\varphi] = \mathcal{F}[T;j[\varphi]] - j[\varphi]\,\varphi \,,\label{Veff2} \ee and
\be \frac{d\,V_{eff}[T;\varphi]}{d\,\varphi} = - j \,.\label{extreV}\ee

This relation is very illuminating, let us first consider it at tree level, without quantum and thermal corrections,  when $V_{eff}[\varphi] = V[\varphi]$. The relation (\ref{extreV}) clearly states that $j$ is the external force necessary to maintain the \emph{space-time constant} $\varphi$ at a value that does \emph{not} correspond to the minimum of the potential. This force vanishes for $\varphi$ satisfying $dV[\varphi]/d\varphi=0$, namely the equilibrium condition. The relation (\ref{extreV}) must be compared to the classical \emph{equation of motion} for a spatially constant (homogeneous) field configuration, namely at tree level   (dots stand for time derivatives)
\be \frac{d}{d\varphi} V[\varphi] = - \ddot{\varphi}\,,\label{eomclas}\ee which when compared with eqn. (\ref{extreV})  clearly states that in \emph{absence of dynamical evolution}, the external force $j$ must be applied to maintain $\varphi$ away from the equilibrium value. This observation will be of paramount importance in the discussion of  dynamics in the next sections.

 As it will become clear in the discussion below, it is more convenient to discuss the effective potential and its dynamical generalization in terms of the \emph{fluctuations} of the field $\phi(\vx)$ around the expectation value $\varphi(\vx)$, a classical c-number field. Hence, as in the zero temperature case, eqn. (\ref{fieldsplit}),  we introduce the field operator
 \be \delta(\vx) = \phi(\vx) - \varphi(\vx)\,,\label{delta} \ee and write
 \be H_J[\phi] \equiv \int d^3 x J(\vx)\,\varphi(\vx) + H_J[\delta] \,, \label{shiftH}\ee where
 \be  {H}_{J}[\delta] \equiv H[\delta+\varphi] + \int d^3 x J(\vx)\,\delta(\vx) \,,\label{Hdel}\ee from which we find
 \be e^{-\beta F[T;J]} = e^{-\beta\,\int d^3 x J(\vx)\,\varphi(\vx)}\, e^{-\beta F_{\delta}[T;J]}\,, \label{fdel} \ee
 with the Helmholtz free energy for the fluctuations around the mean field,
 \be F_{\delta}[T;J] = -\frac{1}{\beta} \,\ln\Big[ \mathrm{Tr}\,e^{-\beta  {H}_J[\delta]}\Big]\,,\label{fdel2} \ee   and
 \be F[T;J] = F_{\delta}[T;J] + \int d^3 x J(\vx)\varphi(\vx)\,.  \label{relafs}\ee With this transformation, the relation (\ref{jfi}) yielding  the source $J(\vx)$ in terms of
 the expectation value $\varphi(\vx)$ is obtained from the \emph{constraint}
 \be \langle \delta(\vx) \rangle =  \frac{\mathrm{Tr} \,\delta(\vx)\, e^{-\beta H_J[\delta]}}{\mathrm{Tr}\, e^{-\beta H_J[\delta]}} = 0 \,.\label{constraint} \ee The Legendre transform (\ref{Gibbs}) yields the generalized Gibbs free energy as
 \be \Omega[T;\varphi] = F_\delta[T;J[\varphi]] \,,\label{nugibs}\ee which upon considering a spatially constant expectation value $\varphi$ yields the effective potential
 \be V_{eff}[T;\varphi] = \mathcal{F}_{\delta}[T,\varphi,j[\varphi]] \,.\label{Veffin}\ee Namely, the effective potential is the Helmholtz free energy density for the   quantum fluctuations around a \emph{space-time constant expectation value}, with the constraint $\langle \delta(\vx)\rangle =0$ which \emph{defines} $j\equiv j[\varphi]$. The condition (\ref{extreV}) yielding
 \be \frac{d\,V_{eff}[T;\varphi]}{d\,\varphi} = - j[\varphi] \,.\label{extreV2}\ee  is now a consistency condition.  This is a main conclusion of this analysis, and while it is  a direct result of the analysis in  the pioneering work in references\cite{jackiw,weinberg}, we emphasize it here because \textbf{i:)} it is an \emph{exact} result, valid to all orders in couplings and loop expansion, \textbf{ii:)} it has important \emph{thermodynamic consequences}, in particular
 \be V_{eff}[T,\varphi] \equiv \mathcal{F}_{\delta}[T,\varphi] = \mathcal{U}[T,\varphi]-T\,\mathcal{S}[T,\varphi] \,\label{US} \ee where
 \be \mathcal{U}[T,\varphi] = \frac{1}{\mathcal{V}}\,\frac{\mathrm{Tr} \, H_J[\delta]\, e^{-\beta H_J[\delta]}}{\mathrm{Tr}\, e^{-\beta H_J[\delta]}} = \frac{\partial}{\partial\beta}\Big\{\beta  \mathcal{F}_{\delta}[T,\varphi]\Big\} \,,\label{Uinten}\ee is the internal energy density, and $\mathcal{S}[T,\varphi]$ is the thermodynamic entropy density as a function of $\varphi$, which by the relations (\ref{US},\ref{Uinten}) is given by\footnote{The partial derivatives are at constant $\varphi$.}
 \be \mathcal{S}[T,\varphi] = - \frac{\partial}{\partial T}  V_{eff}[T,\varphi] \,.\label{entropyV}\ee These are  non-perturbative, exact relations that  link directly the effective potential to the thermodynamic internal energy and entropy. In particular, the relation (\ref{entropyV}) is very important    because when $V_{eff}[T,\varphi]$ is used in a \emph{dynamical} equation of motion for the mean field $\varphi$, its time evolution translates into a time evolution of the entropy density, which must be compatible with the fundamental principles of thermodynamics.

  Under reversible transformations, namely local thermodynamic equilibrium, the entropy obeys the second law of thermodynamics, it remains constant or increases monotonically. This fundamental aspect will be shown below to be in striking contradiction with the use of the effective potential in dynamical situations.

 \subsection{The one-loop effective potential:}\label{subsec1lup}

 With the Hamiltonian (\ref{Hamiltonian1}) we find
\be H_J[\delta] = \mathcal{V}\,V[\varphi]+\int d^3 x \Bigg\{ \frac{1}{2} \pi^2_{\delta}(\vx)+ \frac{1}{2}(\nabla \delta(\vx))^2 + \frac{1}{2}\,V^{''}[\varphi] \,\delta^2(\vx)+\frac{1}{3!}\,V^{'''}[\varphi]\,\delta^3(\vx) + \cdots + \Big(J(\vx)+V^{'}[\varphi]\Big)\,\delta(\vx)    \Bigg\}\,,\label{hamdelta} \ee were primes stand for derivatives with respect to $\varphi$. Neglecting the terms $\propto \delta^3;\delta^4;\cdots$, which yield higher loop corrections,  the constraint $\langle \delta(\vx) \rangle =0$ is fulfilled by setting $J(\vx) \equiv j$, namely a spatial constant, given by
\be j = - V^{'}[\varphi] \,,\label{jzero} \ee thereby cancelling the linear term in $\delta$ in (\ref{hamdelta}). That the condition (\ref{jzero}) yields $\langle \delta(\vx) \rangle =0$  when  neglecting the cubic and higher powers of $\delta$ is clear, because under these conditions  the Hamiltonian is quadratic in $\delta$,     describing a simple free field of squared mass $V^{''}[\varphi]$ for which the density matrix $e^{-\beta H_J[\delta]}$  is Gaussian with zero mean.

 Quantizing the fluctuations by expanding the fluctuation field $\delta(\vx,t)$ as in equation (\ref{quandelta})  with the frequencies
 \be \omega_k(\varphi)=\sqrt{k^2+V^{''}(\varphi)}\,,\label{omegas}\ee  implementing the constraint (\ref{jzero}), and keeping solely the quadratic terms in $\delta$ in (\ref{hamdelta}) yields
\be H_J[\delta] = \mathcal{V}\,V[\varphi]+\sum_{\vk} \hbar \omega_k(\varphi) \big[a^\dagger_{\vk}\,a_{\vk} + \frac{1}{2} \big]\,.\label{Hjay}\ee The calculation of the Helmholtz free energy now becomes a simple textbook exercise in quantum statistical mechanics, the partition function
\be \mathcal{Z} = e^{-\beta F[\varphi]} = e^{-\beta \mathcal{V}\,V[\varphi]}\, \mathrm{Tr}\,\,\Pi_{\vk}\,  e^{-\beta H_{\vk}}\,,\label{Zet}\ee  with
\be H_{\vk} =  \hbar \omega_k(\varphi) \big[a^\dagger_{\vk}\,a_{\vk} + \frac{1}{2} \big]\,. \label{Hks}\ee The trace is calculated in the occupation number basis yielding
\bea \mathrm{Tr}\,\,\Pi_{\vk} \, e^{-\beta H_{\vk}} & = &   \Pi_{\vk} \Bigg[ e^{-\beta \hbar \omega_k(\varphi)/2}\,\sum_{n_{\vk}=0}^{\infty} e^{-\beta \hbar \omega_k(\varphi) n_{\vk}}\Bigg]= \Pi_{\vk} \Bigg[\frac{e^{-\beta \hbar \omega_k(\varphi)/2}}{1-e^{-\beta \hbar \omega_k(\varphi)}} \Bigg] \nonumber \\ & = &  e^{-\frac{\hbar \beta}{2} \sum_{\vk}\omega_k(\varphi)}\,e^{-\sum_{\vk} \ln\Big[ 1- e^{-\beta  \hbar\omega_k(\varphi)}\Big] }\,,\label{traza}\eea passing to the infinite volume limit with
\be \sum_{\vk } \rightarrow \mathcal{V}\,\int \frac{d^3k}{(2\pi)^3} \,,\label{infiV}\ee we find the one loop finite temperature effective potential (\ref{Veffin})
\be V^{(1)}_{eff}[\varphi] = V[\varphi] + \frac{\hbar}{2}\,\int \frac{d^3k}{(2\pi)^3}\,\omega_k(\varphi) + T \,\int \frac{d^3k}{(2\pi)^3} \, \ln\Big[ 1- e^{-\beta  \hbar \omega_k(\varphi)}\Big]\,, \label{Veff1lup} \ee which is the usual result\cite{jackiw,weinberg,kolb,branrmp}. The $T\rightarrow 0$ limit yields the zero temperature one-loop effective potential given by eqn. (\ref{Veff1lupst}), obtained in the previous section via the Hamiltonian approach and the particular coherent state $\ket{\Phi}$  yielding the constraint (\ref{statexvalsdel}). This analysis clearly points out that the coherent state $\ket{\Phi}$ with the constraint (\ref{anni}) is precisely the ground state of the fluctuation Hamiltonian, because in the zero temperature limit, only the ground state contributes to the partition function $\mathcal{Z}$.

An alternative that will prove useful to obtain the equation of motion for the condensate to study dynamics in the next section is to obtain $dV_{eff}/d\varphi$ from equation (\ref{extreV2}) where $j$ is determined from solving the constraint $\langle \delta (\vx) \rangle =0$.  For example, to zeroth order in the loop expansion $V_{eff}[\varphi] = V[\varphi]$ and the tree level condition (\ref{jzero}) satisfies eqn. (\ref{extreV2}). To generate a loop expansion for $j$ we follow ref.\cite{weinberg} and write
\be j= -V'[\varphi] + \hbar j_1 + \hbar^{3/2} j_2 + \cdots \label{jexp} \ee where $j_1,j_2\cdots$ are of $\mathcal{O}(\hbar^0)$, with this expansion and the field expansion (\ref{quandelta}) (showing that $\delta \propto \hbar^{1/2}$) it follows that
\be j_1 \hbar\,\delta \propto \hbar^{3/2} ~~;~~  j_2 \hbar^{3/2}\,\delta \propto \hbar^2  \cdots \,,\label{expah}\ee which are of the same order in $\hbar$ (loop expansion) as $\delta^3\simeq \hbar^{3/2}; \delta^4 \simeq \hbar^{2}\cdots$.

To generate the loop expansion in a systematic manner, we write the Hamiltonian (\ref{hamdelta}) as
\be H_J[\delta]= \mathcal{V}\,V[\varphi]+ H_0 + H_I \,,\label{Hsplit}\ee with
\be H_0 = \int d^3 x \Bigg\{ \frac{1}{2} \pi^2_{\delta}(\vx)+ \frac{1}{2}(\nabla \delta(\vx))^2 + \frac{1}{2}\,V^{''}[\varphi] \,\delta^2(\vx) \,\Bigg\}\,,\label{Hzero} \ee and
\be H_I = \int d^3 x \Bigg\{ \big(J(\vx)+V^{'}[\varphi]\Big)\,\delta(\vx) +  \frac{1}{3!}\,V^{'''}[\varphi]\,\delta^3(\vx) + \frac{V^{''''}[\varphi]}{4!} \,\delta^4(\vx) + \cdots  \Bigg\}\,.\label{HI} \ee
Let us define
\be U(\tau) = e^{\frac{\tau}{\hbar}H_0}\, e^{-\frac{\tau}{\hbar}(H_0+H_I)}\,,\label{Uoft}\ee from which it follows that
\be e^{-\beta (H_0+H_I)} = e^{-\beta H_0} \, U(\hbar \beta) \,.\label{Ubetadef} \ee
$U(\tau)$
obeys the differential equation
\be \frac{d \, U(\tau)}{d\tau} = - \frac{1}{\hbar}\,H_I(\tau) \,U(\tau)~~;~~ U(0)=1 \,, \label{diffU}\ee  where
\be H_I(\tau) = e^{\frac{\tau}{\hbar} H_0}\, H_I \, e^{-\frac{\tau}{\hbar} H_0}\,.\label{HItau}\ee The solution of (\ref{diffU}) is
\be U(\tau) = 1- \frac{1}{\hbar}\int^{\tau}_0 H_I(\tau_1) \,d\tau_1 + \frac{1}{\hbar^2}\int^{\tau}_0\, d\tau_1 \int^{\tau_1}_0  H_I(\tau_1)H_I(\tau_2) \,d\tau_2 + \cdots = T_{\tau}\,\Big(e^{-\frac{1}{\hbar}\int^{\tau}_0 H_I(\tau_1)\,d\tau_1}\Big)\,,\label{soluU} \ee where $T_{\tau}$ is the $\tau$ ordering symbol. Therefore, $U(\tau)$ is the time evolution operator in the interaction picture in imaginary time, namely in the Matsubara representation\cite{wale}.  We can now write the partition function as
\be  \mathrm{Tr}\,e^{-\beta  {H}_J[\delta]} = e^{-\beta \mathcal{V} \mathcal{F}^{(1)}[\varphi]} \, \langle U(\hbar\beta) \rangle_0 \,,\label{partfu}  \ee  where
\be e^{-\beta \mathcal{V} \mathcal{F}^{(1)}[\varphi]} = e^{-\beta \mathcal{V} V[\varphi]}\,  \mathrm{Tr}\,e^{-\beta  {H}_0}  \,,\label{Funo}\ee and the expectation value in the free field theory, defined as
\be  \langle U(\hbar\beta) \rangle_0  = \frac{ \mathrm{Tr}\,e^{-\beta  H_0}\,U(\hbar\beta)}{ \mathrm{Tr}\,e^{-\beta  H_0}}\,,\label{aveU}\ee can be   obtained in a systematic loop expansion. The trace in eqn. (\ref{Funo}) is precisely given by eqn. (\ref{traza}) with the result that $\mathcal{F}^{(1)}[\varphi]$ is the one-loop finite temperature effective potential $ V^{(1)}_{eff}[\varphi]$ given by eqn. (\ref{Veff1lup}). Therefore, from eqn. (\ref{Veffin}) we find the general form of the effective potential
\be V_{eff}[\varphi] = V^{(1)}_{eff}[\varphi] - \frac{1}{\beta \mathcal{V}}\,\ln\Big[\langle U(\hbar\beta) \rangle_0 \Big] \,. \label{fulVeff}\ee

This is an exact result where $j$ is determined by the constraint $\langle \delta(\vx,t)\rangle=0$ order by order in the loop expansion.

We note that $\ln\Big[\langle U(\hbar\beta) \rangle_0\Big]$ begins at two loops, namely $\mathcal{O}(\hbar^2)$ because   $\langle \delta^m(\vx,\tau) \rangle_0 =0$ for odd values of m.

 The expectation value (\ref{aveU}) can now be obtained in a loop  expansion, with the field $\delta$ in the Matsubara interaction picture $\delta(\vx,\tau)=   e^{\frac{\tau}{\hbar} H_0} \delta(\vx,0) e^{-\frac{\tau}{\hbar} H_0}$, namely
 \be \delta(\vx,\tau) =  \frac{\sqrt{\hbar}}{\sqrt{\mathcal{V}}}\,\sum_{\vk} \frac{1}{\sqrt{2\omega_k}}\,\Big[ a_{\vk}\,e^{- \omega_k \tau}\,e^{i\vk\cdot \vx} + a^\dagger_{\vk} \,e^{ \omega_k \tau}\,e^{-i\vk\cdot \vx}\Big]\,. \label{delmatsu} \ee

However, we need the explicit expression for the source $J(x)$, which is determined from the constraint
\be \frac{ \mathrm{Tr}\,e^{-\beta  H_0}\,U(\hbar\beta)\,\delta(\vec{0},0)}{ \mathrm{Tr}\,e^{-\beta  H_0}\,U(\hbar\beta)} = \frac{ \langle U(\hbar\beta)\,\delta(\vec{0},0) \rangle_0 }{ \langle U(\hbar\beta) \rangle_0 } =0 \,.\label{deleqzero} \ee Up to leading order $\hbar^0$ (neglecting cubic and higher order powers of $\delta$  in $H_I$), the numerator yields
\be -\int d^4x \, \big(J(\vx)+V^{'}[\varphi]\big)\, \langle \delta(\vx,\tau)\,\delta(\vec{0},0)\rangle_0 =0 ~~;~~\int d^4x \equiv \int^{\hbar \beta}_0 d\tau \int d^3x \,,\label{zerothev}\ee since $\langle \delta(\vx,\tau)\,\delta(\vec{0},0)\rangle_0\neq 0$ it follows that $J(\vx) = j = -V'[\varphi]$,  which is precisely the relation (\ref{jzero}) with $j$ defined by eqn. (\ref{defs}), and consistent with the expansion (\ref{jexp}).

Now we obtain the $\mathcal{O}(\hbar)$ contribution to $j$ by considering the cubic term in the interaction, yielding
\be -\int d^4x \, \Bigg\{ \big(J(\vx)+V^{'}[\varphi]\big)\, \langle \delta(\vx,\tau)\,\delta(\vec{0},0)\rangle_0 + \frac{V^{'''}[\varphi]}{3!}\,\langle \delta^{3}(\vx,\tau)\,\delta(\vec{0},0)\rangle_0 \Bigg\} =0 \,,\label{juno} \ee using Wick's theorem
\be \langle \delta^{3}(\vx,\tau)\,\delta(\vec{0},0)\rangle_0 = 3\,\langle \delta(\vx,\tau)\,\delta(\vec{0},0)\rangle_0 \,\langle \delta^2(\vx,\tau) \rangle_0 \,,\label{wick}\ee the calculation of $\langle \delta^2(\vx,\tau) \rangle_0$ is straightforward using the expansion (\ref{delmatsu}),  leading to the result
\be -\int d^4x \, \langle \delta(\vx,\tau)\,\delta(\vec{0},0)\rangle_0 \,\Bigg\{  J(\vx)+V^{'}[\varphi] \,  + \frac{\hbar}{2}\,  {V^{'''}[\varphi]}\, \int \frac{d^3k}{2\omega_k}\big[1+2 n_k(\varphi)] \Bigg\} =0 ~~;~~ n_k(\varphi) = \frac{1}{e^{\beta \hbar \omega_k(\varphi)}-1}\,,\label{junofini}\ee from which we find
\be j= -\Big[V^{'}[\varphi]   + \frac{\hbar}{2}\,  {V^{'''}[\varphi]} \, \int \frac{d^3k}{2\omega_k}\big[1+2 n_k(\varphi)] \Big] = - \frac{d}{d\varphi} V^{(1)}_{eff}[\varphi]  \,, \label{jotah} \ee thus explicitly confirming the relation (\ref{extreV2}) with $ V^{(1)}_{eff}[\varphi]$ given by (\ref{Veff1lup})  up to order $\hbar$, namely one loop.

This analysis confirms the ``recipe'' to obtain the one-loop effective potential advocated in ref.\cite{jackiw}: write $\phi= \delta + \varphi$ in the Lagrangian density, expand in the fluctuation $\delta$ up to second order and \emph{neglect} the linear term in $\delta$, the resulting Lagrangian density describes a free field theory of a scalar field of mass squared $\mathcal{M}^2 = V^{''}[\varphi]$. The one loop effective potential is the Helmholtz free energy density of this free field theory.

Although alternative functional methods yield the effective potential in a loop expansion, the main purpose of revisiting and complementing the Hamiltonian framework of ref.\cite{weinberg},   confirming the one loop results of refs.\cite{jackiw,weinberg} and explicitly showing that the zero temperature limit   coincides with the effective potential obtained from the expectation value of the Hamiltonian in the state (\ref{anni}) as shown in ref.\cite{veff}, is to highlight the  following aspects:

\begin{itemize}

\item{The finite temperature effective potential is obtained for \emph{space-time constant} expectation values under the assumption of thermal equilibrium. It is identified with the Helmholtz free energy density under a constraint that the expectation value of the fluctuations around a fixed space-time constant mean field vanish. This constraint is implemented by introducing   a Lagrange multiplier via an external   constant source $J$ coupled linearly to the field. The Lagrange multiplier $J$ represents an external force that keeps the expectation value of the field \emph{in equilibrium} away from the minimum of the effective potential. This force vanishes when the expectation value corresponds to the extremum of the effective potential. This is the content of the exact relation (\ref{extreV}). }

\item{The Hamiltonian formulation of both the zero and finite temperature effective potential explicitly shows that the one loop finite temperature effective potential is the Helmholtz free energy of the free field fluctuations $\delta$ around the expectation value $\varphi$.  Up to one loop, this is a free scalar field theory of squared mass $V^{''}[\varphi]$, which is a \emph{space-time constant}, quantized in terms of the usual mode functions $e^{\pm i \omega_k t}$ of  constant frequency $\omega_k(\varphi)= \sqrt{k^2+V^{''}[\varphi]}$. Furthermore, the distribution function (occupation number)  of these quanta is the usual thermal equilibrium Bose-Einstein distribution with   frequency $\omega_k(\varphi)$.
    The zero temperature limit is the ground state expectation value of the free field Hamiltonian associated of these fluctuations. It is precisely the one loop effective potential obtained from the Hamiltonian method in ref.\cite{veff}.}

\item{The finite temperature effective potential being identified with the Helmoltz free energy density has particular thermodynamic significance because it is directly related to the internal energy density and the \emph{entropy} density, $\mathcal{S}= - \partial V_{eff}[T,\varphi]/\partial T$. This relation is \emph{exact} and entails that \emph{if} the effective potential is used in a dynamical equation of motion of the mean field, fundamental thermodynamic properties of the entropy restrict its domain of validity in such equation of motion. }

    \item{The perturbative method implemented to obtain $j[\varphi]$, yielding eqn. (\ref{jotah}) will be seen below to be very similar to the formulation of the equations
    of motion from non-equilibrium quantum field theory. }

While several of these points seem obvious from the results leading up to the final expression of the effective potential (\ref{fulVeff}) with the  one loop result given by eqn. (\ref{Veff1lup}), when uncritically extrapolated to the dynamical case they will lead to conclusions that are at odds with the fundamental tenets of (local) thermodynamic equilibrium.

\end{itemize}

 \section{Dynamics: open quantum system perspective.}\label{sec:dynamics}

 As discussed   in the previous section, the finite temperature effective potential is a \emph{static quantity}, designed to explore the free energy landscape
 in equilibrium at finite temperature as a function of a space-time constant order parameter, namely the expectation value of the scalar field in equilibrium. Yet, it is often used in dynamical situations in an equation of motion for this homogeneous order parameter:
 \be \ddot{\varphi}(t) + \frac{d}{d\varphi}V_{eff}[\varphi(t)] =0 \,.\label{eom1}\ee  In this section, we endeavor to understand if and under what circumstances such an equation of
 motion in terms of $V_{eff}[\varphi]$ is valid.
  We note that an important consequence of using the static effective potential in   the  equation of motion (\ref{eom1}) is that, in this equation  the effective potential only depends on time via the time evolution of $\varphi$, leading to the conserved quantity
 \be \frac{1}{2} (\dot{\varphi})^2 + V_{eff}[\varphi] =\mathcal{E} = \mathrm{constant} \,.\label{Econs}\ee This result is a direct consequence of \emph{assuming} that the   Helmholtz free energy density   depends on time solely via the time evolution of $\varphi$. A consequence of this equation when combined with the exact result (\ref{entropyV}), is that the thermodynamic entropy depends on time via the time dependence of the mean field.

 It has important implications: let us consider the unbroken symmetry case in which the minimum of the effective potential is at $\varphi =0$, and that the initial value  of $\varphi$ corresponds to a large amplitude with $\dot{\varphi}=0$, hence a large value of $\mathcal{E}$. Then  as $\varphi(t)$ rolls down the potential hill $\varphi$  and consequently $V_{eff}[\varphi]$, become small, however, the velocity $\dot{\varphi}$ has to become large, therefore while $V_{eff}$ is small, its  \emph{time derivative}   becomes large, entailing that the Helmholtz free energy and the \emph{entropy}, which have been obtained in \emph{equilibrium} are actually changing  rapidly in time. This behaviour results in  a  contradiction between the assumptions of thermal equilibrium and the validity of the dynamical equation of motion.

 \vspace{1mm}

 \textbf{Local Thermodynamic Equilibrium:?}

  Using $V_{eff}[\varphi]$, a static function, in the dynamical equation of motion (\ref{eom1}) \emph{suggests} that an underlying (albeit unspelled) assumption is that of local thermodynamic equilibrium (LTE). Namely,  that the distribution function $n_k(\varphi)$ which enters in $dV_{eff}/d\varphi$ (see eqn. (\ref{jotah}))  is always the Bose-Einstein distribution function at temperature $T$ with the frequencies $\omega_k(t) = \sqrt{k^2+V^{''}[\varphi(t)]}$ which are  now   time dependent. This implies that the distribution function adjusts to the change in the frequency   on time scales much shorter than that of the evolution of the frequency itself. Underpinning this assumption is the concept of treating the dynamics of $\varphi$ as a \emph{quantum open system}, namely that the scalar field is in contact with other degrees of freedom that constitute a thermal bath, itself in equilibrium at temperature $T$, with which it exchanges energy-momentum via collisional processes. In postulating eqn. (\ref{eom1}) for the dynamics, the bath itself and its interactions with the scalar field are not   specified.

  A consistent justification of the assumption of (LTE) and the applicability of the effective potential as a function of time through the evolution of $\varphi(t)$ and the distribution functions $n_k(\varphi(t);t)$  would imply solving simultaneously the set of Boltzmann equations for the distribution function with a fully specified collisional term from the coupling
  to the bath degrees of freedom, along with the equation of motion for $\varphi(t)$. Undoubtedly implementing such program is a major undertaking and has not yet been attempted, nor is it our objective in this study. Instead we invoke the usual argument\cite{kolb,bernstein,dodelson} of comparing the time scales of collisional relaxation with those from the   dependence of the
distribution function \emph{assuming} the validity of (LTE) in its time evolution. Such arguments are ubiquitous in cosmology and underpin the  understanding of the validity of (LTE) during cosmological expansion as well as the freeze-out of species and decoupling from a thermal environment\cite{kolb,bernstein,dodelson}.

  In absence of external forces, and in a homogeneous situation the distribution function obeys the Boltzmann equation\cite{kolb,bernstein,dodelson}
  \be \frac{d}{dt}n_k(t) = \mathcal{C}[n_k] \,,\label{boltz}\ee where $\mathcal{C}[n_k]$ is the collision kernel. The \emph{exact} distribution function, solution of this Boltzmann equation is written as $n_k(t)=n_{lte}(k;t)+\delta n_k(t)$ where $n_{lte}(k;t)$ is the (LTE) distribution function, and  $\delta n_k(t)$ is the departure from  (LTE). If $\delta n_k(t)/n_{lte}(k;t) \ll 1$ then (LTE) is a reliable approximation to the exact distribution function.  The departure from (LTE) is studied within a Chapman-Enskog expansion\cite{liboff}, in terms of the ratios between the relaxation time and the time scale of variation of the distribution function, and the ratio of the mean free path to the spatial scale of variation (Knudsen number). (LTE) ensues when these ratios are $\ll 1$. In absence of a specific model for the collision kernel we can resort to the relaxation time approximation  for a qualitative (and semi quantitative) estimate\cite{bernstein,kolb,dodelson}\footnote{See section 4, eqn. (4.41) and following in ref.\cite{bernstein}.},
  \be \frac{d}{dt}n_k(t) = -\frac{1}{\mathcal{T}}\,\delta n_k(t) ~~;~~ \delta n_k(t) = (n_k(t)-n_{lte}(k;t)) \,,\label{relap}\ee where $ \mathcal{T}$ is the average time between collisions, i.e.  relaxation time or inverse reaction rate $\Gamma=1/\mathcal{T}$\cite{bernstein,kolb,dodelson,liboff},
  \be \mathcal{T} = \frac{1}{n\langle \sigma v \rangle} \,,\label{reltime}\ee with $n$ the density of scatterers, $\sigma$ the cross section and $v$ the relative velocity.   We take the (LTE) distribution function
  \be n_{lte}(k;t) \equiv \frac{1}{e^{\beta \omega_k(t)}-1} \,,\label{letn}\ee since this is the distribution function that enters in $V_{eff}[\varphi(t)]$. To first order in the Chapman-Enskog expansion\cite{bernstein,liboff}, $n_k(t)$ on the left hand side of eqn. (\ref{relap}) is replaced by $n_{lte}(k;t)$, therefore to this order
    \be \delta n_k(t) = - \mathcal{T} \frac{d}{dt}n_{lte}(k;t)\,,\label{firstce}\ee and (LTE) is a reliable approximation if
    \be \frac{\delta n_k(t)}{n_{lte}(k;t)} \ll 1 \,.\label{letok}\ee  Let us consider the high temperature limit $\beta \omega_k \ll 1$ where we expect a short relaxation time, yielding
    \be \Big| \frac{\delta n_k(t)}{n_{lte}(k;t)}\Big| \simeq \mathcal{T}\,\Big|\frac{\dot{\omega_k}(t)}{\omega_k(t)}\Big|\,, \label{ratiolte}\ee therefore (LTE) in this regime is fulfilled when
    \be \mathcal{T}\,\Big|\frac{\dot{\omega_k}(t)}{\omega_k(t)}\Big|\ \ll 1 \,.\label{lteok} \ee

    Approximating the high temperature limits
    \be n \simeq T^3~;~ \langle \sigma v \rangle \simeq \frac{g^2}{T^2} \,,\label{hiTng}\ee with $g$ a dimensionless coupling, and considering long-wavelength fluctuations, since
    we expect these to be the slowest to relax to (LTE), the condition (\ref{letok}) yields
    \be \Bigg|\frac{V^{'''}[\varphi]\,\dot{\varphi}}{2 g^2 T V^{''}[\varphi]}\Bigg|\ll 1 \,,\label{ltecondi}\ee  which obviously depends not only on the coupling $g$ to the bath
    but also the details of the potential $V[\varphi]$ such as mass and couplings. To be specific,  let us consider the case with tree level potential
\be V[\varphi] = \frac{m^2}{2}\varphi^2 + \frac{\lambda}{4}\varphi^4 ~~;~~ m^2 >0\,,\label{treeV}\ee  with a large amplitude initial condition
\be \dot{\varphi}(0)=0~~;~~  \lambda \varphi^4(0) \simeq T^4~~;~~\lambda \varphi^2(0) \gg m^2\,,\label{inicon}\ee consistently with an initially thermalized state of large energy density $\propto T^4$. As $\varphi(t)$ rolls down the potential hill, well before reaching
the minimum (for example $\varphi(t) \simeq \varphi(0)/2$) it follows that $\dot{\varphi}(t) \simeq \sqrt{\lambda} \varphi^2(0)~;~V^{'''}[\varphi] \simeq \lambda \varphi(0)~;~V^{''}[\varphi] \simeq \lambda \varphi^2(0)$ and the ratio (\ref{ltecondi}) implies that the condition for the validity of (LTE) becomes
\be \frac{ {\lambda}^{1/4}}{g^2} \ll 1 \,,\label{ratiocoups}\ee which may very well be violated depending on a delicate balance, in other words a fine tuning,  of the strengths of couplings.  For example, if the collisional kernel refers to collisions among the quanta of the scalar field, with self-interaction given by the potential (\ref{treeV}) then $g^2 \rightarrow \lambda^2$ with an obvious violation of the (LTE) condition for weak coupling.

While a more careful treatment of the quantum kinetics combined with the equation of motion for $\varphi$ is required for a thorough assessment of the validity or breakdown of (LTE), a program well beyond the scope of this study,  this simple analysis highlights that the validity of (LTE) must be carefully assessed and should not be taken for granted as it may not be fulfilled and may imply fine tuning in generic cases.

The reader will recognize that the criterion for the validity of (LTE) is the usual one invoked in cosmology\cite{bernstein,kolb,dodelson}, where the reaction rate $\Gamma = 1/\mathcal{T}$ is compared to the Hubble expansion rate $H$: in the cosmological  case (LTE) is valid for $H/\Gamma \ll 1$. In the dynamical case under consideration,   $\dot{\omega}_k(t)/ {\omega}_k(t)$ replaces $H$, however, other than this difference,  which ultimately is a difference on time scales, the main arguments are indeed similar.

   The effective potential being identified with the Helmholtz free energy density, implies that
\be V_{eff}[\varphi] = \mathcal{U}[\varphi]- T \mathcal{S}[\varphi] \,,\label{Sdef} \ee where up to one loop
\be \mathcal{U}= \frac{\langle H_J[\delta] \rangle}{\mathcal{V}}= V[\varphi] + \frac{\hbar}{2}\,\int \frac{d^3k}{(2\pi)^3}\,\omega_k \,[1+2 n_k(\varphi)]\,,\label{Uen}\ee is the internal energy density, and
\be \mathcal{S} = \int \frac{d^3 k}{(2\pi)^3}\,\Big\{(1+n_k(\varphi))\,\ln (1+n_k(\varphi))- n_k(\varphi)\,\ln (n_k(\varphi))\Big\}\,,\label{Seq}\ee is the entropy density, with the occupation numbers
\be n_k(\varphi) = \frac{1}{e^{\beta   \omega_k(\varphi)}-1}\,.\label{nkfis}\ee
  In the dynamical case $\varphi(t)$ depends on time, consequently the occupation numbers $n_k(\varphi)$ do depend on time via the frequencies $\omega_k(\varphi(t))$.  As a result, the entropy density depends on time, and its time derivative  is given by
 \be \dot{\mathcal{S}}= - \frac{V^{'''}[\varphi(t)]\,\dot{\varphi}}{2\,T^2}\,\int \frac{d^3 k}{(2\pi)^3}\,   {n_k(\varphi)(1+n_k(\varphi))}\,.\label{sdot}  \ee Consider the tree level potential (\ref{treeV}) with $V^{'''}(\varphi(t)) = 6\,\lambda \varphi(t)$, when $\varphi$ is oscillating near the minimum at $\varphi =0$, with $\varphi(t) = \varphi(0)\cos(mt)$, it follows that $\dot{S} \propto \sin(2mt)$. This behavior is actually more general, when the symmetry is unbroken, $\varphi$ oscillates around the minimum and $V^{'''}[\varphi(t)]\,\dot{\varphi}(t)$ changes sign, thereby alternating between increasing   and decreasing entropy along the trajectory. Namely, the entropy density is a \emph{non-monotonic function of time}, a behaviour that, \emph{a priori} is not compatible with a thermodynamic entropy. It may be argued that in the quantum open system approach the entropy of the system may not be a monotonic function of time as the system exchanges energy and momentum with the bath,  and that the change in entropy of the system reflects heat transfer to and from the bath, while the \emph{total} entropy   of the system plus the bath increases monotonically or remains constant. However, we emphasize that the non-monotonicity is in the entropy \emph{density}, therefore the change in entropy is \emph{extensive}, therefor such an argument implicitly accepts that the bath itself is \emph{not} in thermal equilibrium and its dynamics is affected by the system in an extensive manner. Clearly these arguments must be quantified, however, the point remains that the time dependence of the entropy raises relevant questions on the validity of (LTE) in the dynamical evolution of the mean field.

\textbf{Caveats: parametric and spinodal instabilities.}

One of the main objectives of comparing the finite temperature effective potential to the zero temperature effective potential obtained in ref.\cite{veff} is to highlight that the main caveats associated with using the effective potential in the dynamical equation of motion (\ref{eom1})  discussed in this reference also apply to the one-loop finite temperature effective potential (\ref{Veff1lup}). After all taking the $T\rightarrow 0$ limit in this expression yields the one-loop effective potential obtained in ref.\cite{veff} in the Hamiltonian formulation.

The previous analysis on the validity of (LTE), based on a collisional Boltzmann equation,  does \emph{not} include the possibility of instabilities which lead to particle production and non-thermal distribution functions. Two ubiquitous instabilities were studied in detail within the context of the zero temperature effective potential  in ref.\cite{veff}: parametric and spinodal, the latter ones associated with spontaneous symmetry breaking. While we refer the reader to this reference for further details,   for completeness of presentation we summarize here the main aspects of both instabilities, with the objective of emphasizing that both  prevent a formulation of an \emph{equilibrium} finite temperature effective potential as described in the first section.

\begin{figure}[ht!]
\begin{center}
\includegraphics[height=4in,width=3.2in,keepaspectratio=true]{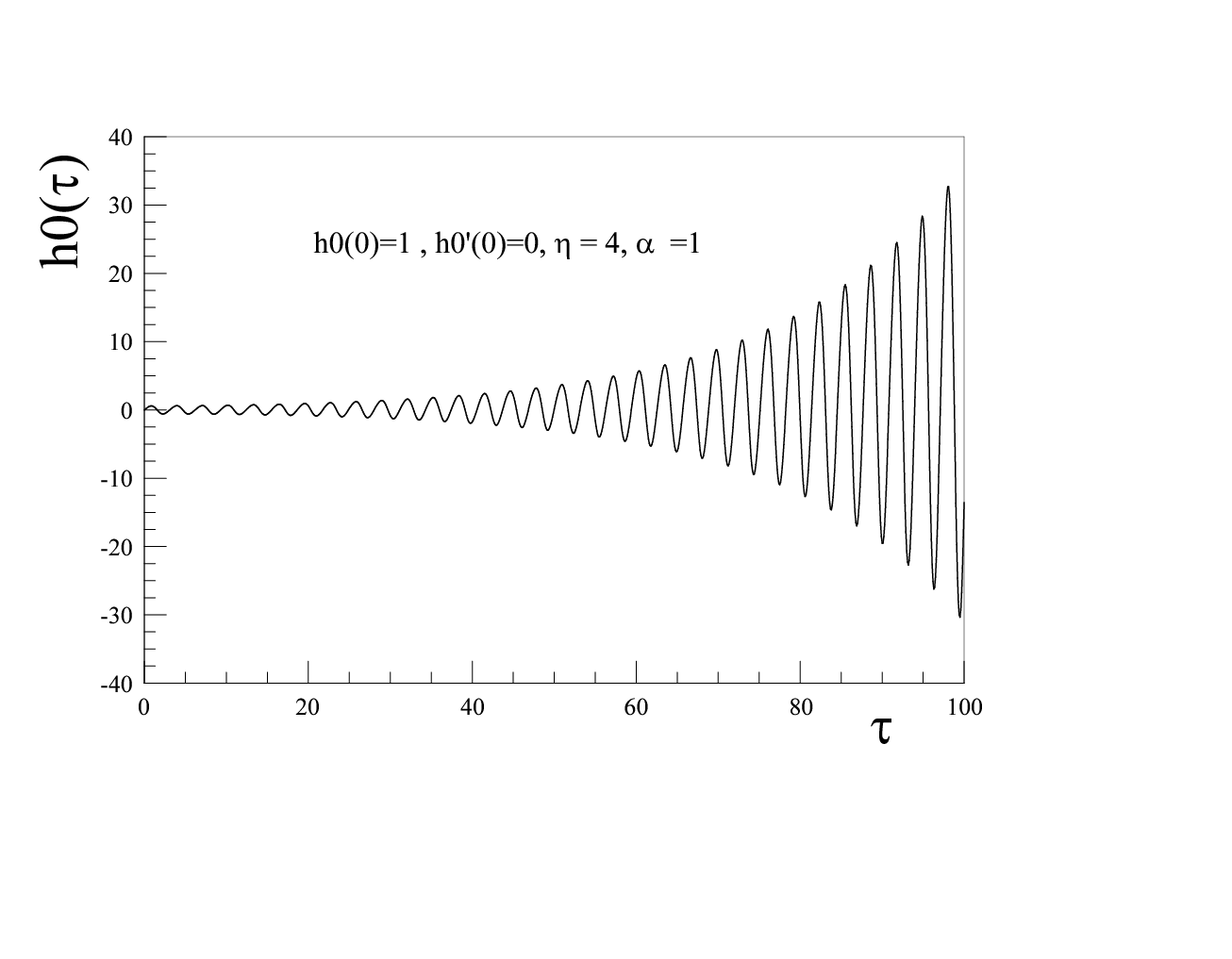}
\includegraphics[height=4in,width=3.2in,keepaspectratio=true]{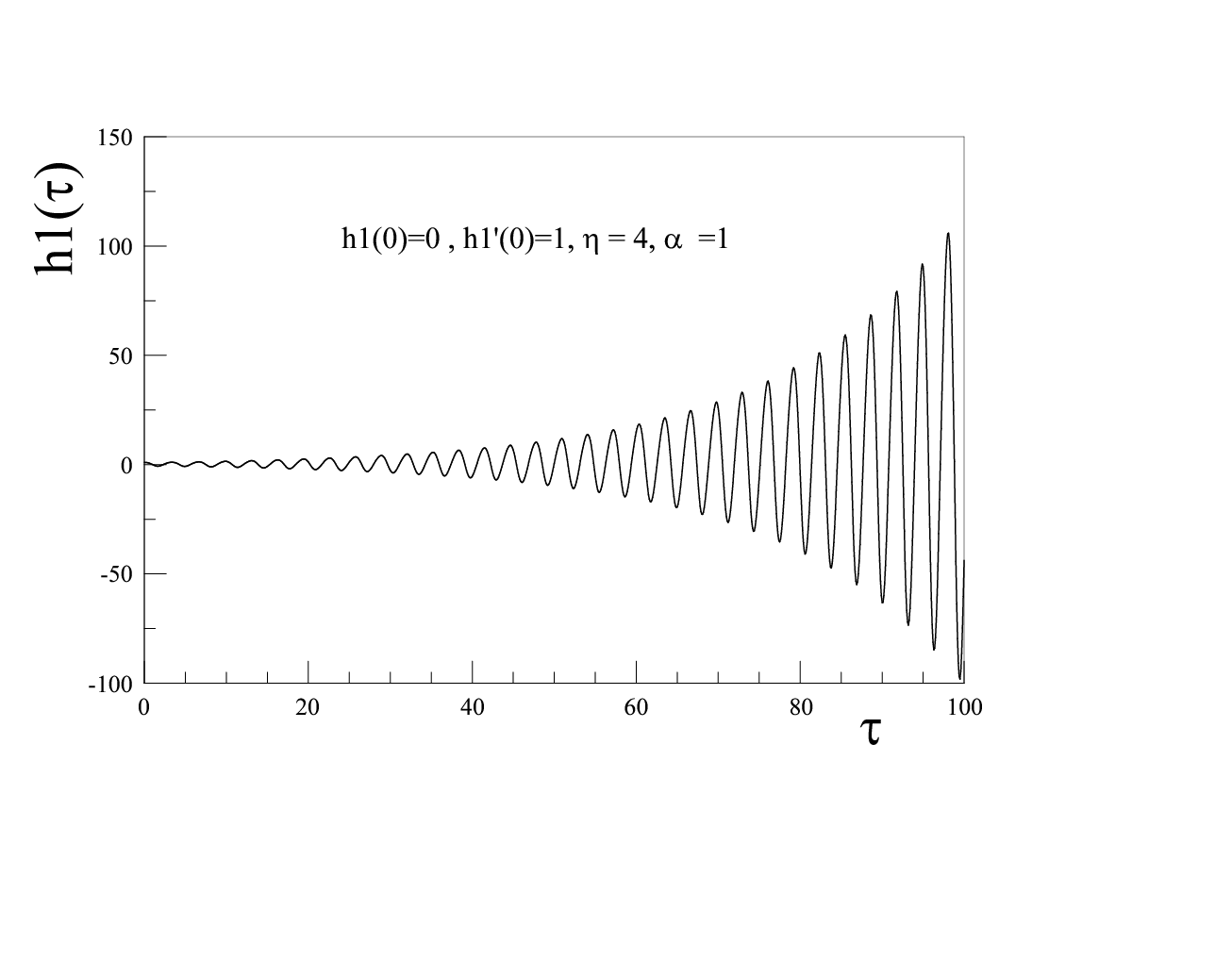}
\caption{Two linearly independent solutions of Mathieu's eqn. (\ref{mathieu}), $h0(\tau),h1(\tau)$ with initial conditions $h0(0)=1,h0'(0)=0;h1(0)=0, h1'(0)=1$, for the first  unstable band. A general solution for a mode function $g_k(\tau)$ is a complex linear combination of $h0(\tau)$ and $h1(\tau)$ satisfying the condition (\ref{wron}). }
\label{fig:lisolns}
\end{center}
\end{figure}

 Let us first consider the tree level potential (\ref{treeV}), yielding $V^{''}[\varphi(t)] = m^2 + 3 \lambda \varphi^2(t)$ with $m^2 >0$ and small amplitude oscillations around the minimum at $\varphi=0$, namely
\be \varphi(t) = \varphi(0)\,\cos(mt) \,.\label{smallos} \ee  yielding
\be V^{''}(\varphi(t))= m^2 + 3 \lambda  \varphi^2(0)\,\cos^2(mt)\,.\label{vsecoft}\ee
Quantization of the fluctuation field $\delta$\cite{veff} with an effective mass squared $V^{''}[\varphi(t)]$ given by (\ref{vsecoft}) leads to Mathieu's equation\cite{mathieu1,abra,kova}, which features instability bands from parametric amplification describing profuse particle production\cite{veff,rehe1}-\cite{rehe6}. While we refer the reader to ref.\cite{veff} and references therein for a more detailed discussion,   for consistency and completeness of presentation  we summarize here some of the important aspects of parametric instability in this case. Introducing the dimensionless variables
\be \tau = mt-\frac{\pi}{2}~~;~~  \alpha = 3\lambda \frac{\varphi^2(0)}{4\,m^2}~~;~~ \eta_k = 1+ \kappa^2+2\alpha ~~;~~ \kappa = \frac{k}{m}\,, \label{matvars}\ee the mode equations (\ref{modeg}) become of the form of Mathieu's equation\cite{mathieu1,abra,kova}
\be \frac{d^2}{d\tau^2}\,g_k(\tau)+ \big[\eta_k- 2\alpha\,\cos(2\tau)\big]g_k(\tau) =0 \,,\label{mathieu} \ee whose instability bands have been analyzed in refs.\cite{mathieu1,abra,kova,veff}. Figure (\ref{fig:lisolns}) displays  two linearly independent solutions in the first instability band, showing the exponential growth from parametric instability.

 A general solution $g_k(t)$ is a linear combination of these two linearly independent solutions. The width of each band, labelled by an integer index $n=2,3 \cdots$  is found to be proportional to $\alpha^n$\cite{veff}.
Within these instability bands the amplitudes of the mode functions grow as $g_k(t) \propto e^{\nu_k t}$ with real $\nu_k \geq 0$   being the Floquet exponents, the smaller wavevectors feature the   largest $\nu_k$ and wider  bandwidth of the unstable regions\cite{veff}.

These instabilities and the concomitant particle production  clearly indicate that   maintaining (LTE) by collisional coupling to a bath is   not a warranted assumption and in general implies fine tuning of couplings to the bath. Furthermore, particle production in the parametrically unstable bands results in non-thermal distribution functions which cannot be approximated by the usual Bose-Einstein distribution functions that emerge in the equilibrium description because particle production is effective within localized bands in momentum. If collisional processes distribute the particles outside the unstable bands into an (LTE) Bose-Einstein distribution function, such processes must occur on time scales shorter than the inverse of the largest Floquet exponent, again implying a fine tuning between the coupling to the heat bath and the parameters of the potential. Clearly this is not a generic situation and depends on particular models and couplings to the bath degrees of freedom.

Let us now consider the case in which the tree level potential leads to spontaneous symmetry breaking, for example the potential (\ref{treeV}), but with $m^2= -\mu^2$ with $\mu^2 >0$, yielding $V^{''}[\varphi] = -\mu^2 + 3 \lambda \varphi^2$. Within the (classical) spinodal region $\varphi^2 < \mu^2 /3$ it follows that $V^{''}[\varphi]< 0$ and there is a band of spinodally unstable wavevectors $k^2 < |V^{''}[\varphi]|$ for which the mode functions grow exponentially $g_k(t) = s_k \,e^{\int^t \nu_k(t')dt'}+ r_k \,e^{-\int^t \nu_k(t')dt'} $ with $\nu_k(t) = \sqrt{|V^{''}[\varphi(t)]|-k^2}>0 $, and $s_k,r_k$ determined by initial conditions, thereby   signalling exponential growth of fluctuations. In this case, for $|\varphi|$ within the spinodal region and values of $k$ in the unstable band, the frequencies $\omega_k[\varphi]= \sqrt{k^2+V^{''}[\varphi]}$ are \emph{purely imaginary} and the Helmholtz free energy density, namely the effective potential,   the internal energy density, occupation numbers $n_k(\varphi)$  and \emph{entropy} (\ref{Seq}) are all \emph{complex}, an untenable situation from the thermodynamic perspective, even when $\varphi(t)$ ``rolls down'' the potential hill very slowly within the spinodal. That equilibrium thermodynamics (or (LTE)) \emph{cannot} describe this situation is well known in statistical physics: the spinodal instabilities are associated with the dynamical process of phase separation and the growth of correlated ordered domains\cite{langer1}-\cite{allen}, and has also been studied in quantum field theory\cite{weinbergwu,calzetta,boyaspino}. Particle production in these spinodally unstable bands is followed by particle production by parametric instabilities when the mean field $\varphi$ has rolled down below the spinodal point and oscillates near the (broken symmetry) minimum, again resulting in non-thermal distribution functions as a result of particle production from parametric amplification in the unstable bands.

Both types of instabilities  result in profuse particle production and non-thermal distribution functions for the produced particles which are localized in momentum within the unstable bands. A redistribution of particles into thermal distribution functions, the underlying assumption in using the finite temperature effective potential,  implies a strong coupling to a thermal bath in such a way that this redistribution occurs on time scales much shorter than the time scales associated with the instabilities. Obviously   if and when such a coupling arises is a model dependent, highly fine tuned and non-generic case.

\textbf{Partial summary:}

In the previous sections we have shown that the finite temperature effective potential is associated with the  \emph{equilibrium} Helmholtz free energy density as a function of the mean field $\varphi$. This is an \emph{exact result} valid to all orders in couplings and loop expansion  with important thermodynamic implications,  and assessed  whether using the effective potential in the equation of motion for $\varphi$ is warranted.

Based on the following aspects, our conclusions are that the regime of validity of the \emph{static effective potential} to describe the dynamics of the mean field is very limited.

\begin{itemize}
\item{Using $V_{eff}(\varphi(t))$ in the equation of motion with $V_{eff}[\varphi]$ obtained in equilibrium quantum field theory, \emph{assumes} that there is (LTE), at a fixed
constant temperature, presummably maintained via a coupling to a thermal bath in equilibrium at such temperature. Although the coupling to the thermal bath is in general not specified, we have provided general arguments based on the Boltzmann equation with a collision term in the relaxation time approximation to suggest that (LTE)   is not warranted in many relevant cases, unless there is a fine tuning  of couplings to the thermal bath.  }

\item{ We have found severe caveats in the cases both without and with symmetry breaking tree level potentials. In absence of symmetry breaking, parametric instabilities associated with the oscillatory dynamics of the mean field near the minimum of the potential, leads to profuse particle production, with distribution functions that are not thermal, and more importantly, a \emph{non-monotonic} behavior of the entropy. While this latter behavior may be argued to describe an exchange of entropy with an external bath, it runs counter to the main tenets of local equilibrium thermodynamics. }

    \item{In the case when the tree level potential admits broken symmetry minima, spinodal instabilities prevent an (LTE) description of the dynamics. In particular for a band of spinodaly unstable wavevectors when $\varphi$ is within the classical spinodal region within which $V^{''}(\varphi) < 0$, the effective frequencies are purely imaginary,  fluctuations grow exponentially, yielding a \emph{complex} effective potential, distribution function $n_k(\varphi)$, internal energy and \emph{entropy}. Whereas the imaginary part of the  internal energy may be associated with a decay rate of a particular non-equilibrium state\cite{weinbergwu,boyaspino}, an imaginary part of the entropy  is untenable, and unacceptable in   thermodynamics.  }

        \item{The caveats emerging from assuming that the effective potential can be used in the dynamical evolution of the mean field, cannot be overcome by any resummation program in \emph{equilibrium} quantum field theory, such as, for example resummation of ``hard-thermal loops''\cite{htl}. Such non-perturbative resummation frameworks cannot possibly address the \emph{dynamical} instabilities associated with parametric amplification or spinodal decomposition, the latter being a hallmark of the early stages of phase separation, the formation and growth of correlated domains and coarsening during a phase transition\cite{langer1}-\cite{allen}. }

\end{itemize}

 \section{Dynamics: decoupling and  freeze out,   closed quantum system evolution.}\label{sec:closed}
 The discussion in the previous section outlines several problems inherent in merely using the finite temperature effective potential to describe the behavior of a dynamical expectation value/condensate. Critically, employing this effective potential tacitly assumes a persistent local thermodynamic equilibrium between the condensate and the environment which requires a precise analysis of the couplings to the thermal bath.  This obfuscates the problem and prevents one from making simple, model-independent statements about the dynamics of the condensate under such conditions.
However, one may consider a closely related scenario wherein a condensate, which was previously in local thermodynamic equilibrium, decouples from the bath and proceeds to evolve in time. In this section we will investigate these dynamics, thereby providing an avenue for studying the behavior of the mean field beyond the time scale when (LTE) is no longer warranted. This problem is not only both tractable and relevant in its own right, but it will provide a useful comparison to the phenomenologically motivated approach of using the effective potential in the equation of motion (see equation (III.1)).

 Let us first consider the case when the tree level potential does not feature spontaneous symmetry breaking, and an initial condition on the mean field such that its velocity is very small and it is up the potential hill, far from the minimum of the tree level potential so that it does not feature oscillations that lead to parametric amplification and breakdown of (LTE) in generic cases. As $\varphi$ rolls down the potential hill with a small initial velocity, there is a time interval when the evolution of the mean field is slow and the condition (\ref{lteok}), or alternatively (\ref{ltecondi}),  for the validity of  (LTE) is fulfilled. This entails that the instantaneous frequencies $\omega_k(\varphi(t))$ are varying slowly on the relaxation time scale, this can be quantified in an adiabatic expansion of the solutions of the mode equations (\ref{modeg})\cite{veff}. Proposing the Wentzel-Kramers-Brillouin (WKB) solution
     \be g_{k}(t) = \frac{e^{-i\int_{t_0}^t\,\Omega_k(t')dt'}}{\sqrt{2\Omega_k(t)}}\,,\label{adiadef} \ee
which when inserted into equation (\ref{modeg}) reveals that $\Omega_k(t)$ must satisfy
    \be \Omega^2_k(t) = \omega^2_k(t)-\frac{1}{2}\Bigg[ \frac{\ddot{\Omega}_k}{\Omega_k} -\frac{3}{2}\frac{\dot{\Omega}_k^2}{\Omega_k^2}\Bigg]\,. \ee
The resulting equation can be solved in an \emph{adiabatic expansion}
    \be \Omega_k^2(t) = \omega_k^2(t) \Bigg[1-\frac{1}{2}\frac{\ddot{\omega}_k}{\omega_k^3}+\frac{3}{4}\Big(\frac{\dot{\omega}_k}{\omega_k^2}\Big)^2+\cdots \Bigg]\,. \label{adiaexp1}\ee Assuming a slow initial evolution, let us consider the leading  (zeroth)  adiabatic order, namely
    \be g^{(0)}_{k}(t) = \frac{e^{-i\int_{t_0}^t\,\omega_k(t')dt'}}{\sqrt{2\omega_k(t)}}\,,\label{zeroadia} \ee as the mean field evolves, its velocity   increases, and at some time scale $t_0$ (LTE) breaks down and the system can no longer remain in thermal contact with the bath. This is the physics of decoupling between the system and the bath. From this time scale onwards, the scalar field evolves independently of the bath, this situation is similar to the decoupling of photons in cosmology, when the mean free path from Thompson scattering is larger than the Hubble radius, the photons evolve freely. Within this context, the time of decoupling is referred to as the ``surface of last scattering'' and is often approximated to be an instantaneous process.

    We model the similar situation as an ``instantaneous'' decoupling assuming that the density matrix for the fluctuations around the mean field is \emph{frozen} in the Heisenberg picture, to that describing free field fluctuations with the frequencies at the decoupling time $t_0$. This assumption leads to the following initial conditions  to leading adiabatic order from  eqn. (\ref{zeroadia}):
    \be g_k(t_0) = \frac{1}{\sqrt{2\omega_k(t_0)}}~~;~~ \dot{g}_k(t_0) = -i   \frac{\omega_k(t_0)}{\sqrt{2\omega_k(t_0)}}~~;~~ \omega_k(t_0)=\sqrt{k^2+V^{''}(\varphi(t_0))}\,,\label{inigis}\ee where $\omega_k(t_0)$ are real and  positive under the assumption that the mean field is within a region with $V^{''}(\varphi(t_0))>0$, which is always the case for potentials with unbroken symmetry.  The initial conditions (\ref{inigis}) imply that the Wronskian condition (\ref{wron}) on the mode functions $g_k(t)$ is satisfied.

    Assuming the validity of (LTE) up to the decoupling time $t_0$, the initial density matrix at this time is taken to be given by
     \be \rho(t_0) = \frac{e^{-\beta_0 H_0}}{\mathrm{Tr}e^{-\beta_0 H_0} }\,,\label{rhozeroin1}\ee
     where
     \be H_0 \equiv \sum_{\vk} \hbar\, \omega_k \Big(a^\dagger_{\vk} \, a_{\vk} + \frac{1}{2}\Big)\,.\label{Hzerodyn1}\ee This choice of initial density matrix is consistent with
     the one-loop effective potential, which is determined by a density matrix describing a free field with a squared mass $V^{''}(\varphi)>0$ as discussed in section (\ref{subsec:finiteT}), in particular the fluctuation Hamiltonian in eqn. (\ref{Hjay}).

When the tree level potential features broken symmetry minima and a spinodal region wherein $V^{''}(\varphi) < 0$, the situation is much more subtle. The band of wavevectors $0 \leq k < |V^{''}(\varphi)|$ is spinodally unstable, (LTE) is not fulfilled regardless of the value of $\dot{\varphi}$ and as discussed in the previous section, the mode functions within this band feature (nearly) exponential growth in time as a consequence of the instability.   The initial conditions (\ref{inigis}) are valid  if $V^{''}(\varphi)>0$, in other words $\varphi$ is \emph{outside} the spinodal region,   however they  must be modified if $\varphi$ is within the unstable region where $V^{''}(\varphi)< 0$.  Nevertheless, we can \emph{parametrize} the initial conditions on the mode functions within the unstable band at some initial time $t_0$ as
\be g_k(t_0) = \frac{1}{\sqrt{2W_k}}~~;~~ \dot{g}_k(t_0) = -i   \frac{W_k}{\sqrt{2W_k}}~~;~~ W_k = \sqrt{k^2+\mathcal{M}^2}~~;~~\mathcal{M}^2 >0 \,.\label{inigispino}\ee which again imply that the Wronskian condition is fulfilled. The effective mass term $\mathcal{M}^2>0$ is a parametrization of the initial condition at a time $t_0$, its actual value depends on the precise ``misalignment'' mechanism that has resulted in the mean field $\varphi$ to be within the spinodally unstable region and must be specified for particular realization of the dynamics.

\textbf{An explicit example: a ``quenched'' phase transition:}
Let us consider the   case of a rapid phase transition modelled by a scalar field theory with a time dependent mass term with Lagrangian density
\be \mathcal{L} = \frac{1}{2} \partial^\mu \phi \partial_\mu \phi + \frac{a }{2} \Big(T^2(t)-T^2_c\Big) \phi^2 + \frac{\lambda}{4} \phi^4 \,,\label{ptlag}\ee with $a>0$ a dimensionless constant,  and a time dependent temperature
\be T(t) = T_i\Theta(t_0-t)+T_f\Theta(t-t_0)~~;~~ T_i > T_c~~;~~T_f<T_c \,. \label{tempoft}\ee This situation describes a sudden phase transition at time $t=t_0$ from an unbroken symmetry case for $t<t_0$ with $T_i > T_c$ to a broken symmetry case for $t>t_0$ with $T_f<T_c$. If for $t<t_0$ the mean field $\varphi$ is oscillating with small amplitude around the equilibrium minimum of the potential at $\varphi=0$ (for $t<t_0$),  and  at the transition time $t_0$ is found with a value $\varphi_0$, the mode functions are of the form $e^{\pm iW_k t}$ with
\be W_k=\sqrt{k^2+\mathcal{M}^2}~~;~~ \mathcal{M}^2= a(T^2_i-T^2_c)+3\lambda \varphi^2_0 >0\,.\label{Mpt}\ee

For $t>t_0$, after the temperature dropped to $T_f < T_c$  the mean field is now within the spinodal region if the initial value $\varphi_0$ is such that $a(T^2_f-T^2_c)+3\lambda \varphi^2_0 < 0 $, the mode functions at the transition time have precisely the initial conditions (\ref{inigispino}). After this sudden transition, the mean field will begin  rolling  down the potential hill, and the mode functions $g_k(t)$ that describe the fluctuations will grow nearly exponentially while the mean field is within the spinodal. This simple but relevant example explicitly describes a physical situation in which the mean field is found initially within the spinodal region. The ensuing time evolution of the mode functions exhibit the (nearly) exponential growth associated with the dynamics of the phase transition and the emergence of correlated domains with a growing correlation length\cite{weinbergwu,boyaspino}.

This specific example is by no means exhaustive, nor do we dwell here on the ``quenching mechanism'',  but it highlights that in the dynamical case the ``misalignment'' mechanism by which the initial value of the mean field is found inside the spinodal region must be specified, \emph{along} with the initial conditions on the mode functions that describe the fluctuations around the mean field.

Both cases, with and without spontaneous symmetry breaking in the tree level potential can be summarized by the following initial conditions that satisfy the Wronskian condition
\be g_k(t_0) = \frac{1}{\sqrt{2W_k}}~~;~~\dot{g}_k(t_0) = -i\frac{W_k}{\sqrt{2W_k}}~~;~~ W_k =\sqrt{k^2+M^2}~~;~~ M^2 \equiv  \Bigg\{\begin{array}{c}
                                                                                                                            V^{''}(\varphi(t_0))  ~~;~~V^{''}(\varphi(t_0))>0\\
                                                                                                                            \mathcal{M}^2 >0  ~~;~~V^{''}(\varphi(t_0))<0
                                                                                                                         \end{array}
 \,. \label{sumainiconds}\ee

Under the assumption of instantaneous decoupling and
 to establish a relation with the thermal density matrix discussed in the previous section (\ref{subsec:finiteT}) within the context of the static finite temperature effective potential, let us introduce
 \be H_0 \equiv \sum_{\vk} \hbar\, W_k \Big(a^\dagger_{\vk} \, a_{\vk} + \frac{1}{2}\Big)\,,\label{Hzerodyn}\ee and the initial density matrix is taken to be given by
 \be \rho(t_0) = \frac{e^{-\beta_0 H_0}}{\mathrm{Tr}e^{-\beta_0 H_0} }\,,\label{rhozeroin}\ee
 where the frequencies $W_k$ are given by eqn. (\ref{sumainiconds}) and the time independent annihilation and creation operators are the same that enter in the quantization of the fluctuation field $\delta(\vx,t)$, given by (\ref{delex}).

    This particular choice of the initial density matrix is motivated by an ``instantaneous decoupling'' from (LTE) and has a clear and simple interpretation: it describes a free field theory for the fluctuations with squared mass $M^2>0$ given by eqn. (\ref{sumainiconds})   in thermal equilibrium at a temperature $T_0 = 1/\beta_0$.

    The main assumption behind this choice is that the coupling to the thermal bath maintains (LTE) up to time $t_0$ at which the  time scale of change of the frequencies is much shorter  than the relaxation time and the scalar field decouples instantaneously from the bath. From this time onwards the density matrix follows unitary time evolution determined by the
    dynamics of the scalar field.

 Note the similarity with the fluctuation Hamiltonian, the second term on the right hand side of eqn. (\ref{Hjay}) which yields the static one-loop effective potential, however, unlike the frequencies (\ref{omegas}) that enter in (\ref{Hjay}), which are imaginary within the spinodal region, the $W_k$ that enter in $H_0$  are always real.

  In this (Gaussian) density matrix it follows that
 \bea && \langle  a^\dagger_{\vk}\rangle = \mathrm{Tr}\, a^\dagger_{\vk}  \,\rho(t_0) =0 ~~;~~ \langle  a_{\vk}\rangle = \mathrm{Tr}\, a_{\vk}  \,\rho(t_0) =0 \,,\nonumber \\
&&  \langle a^\dagger_{\vk} a_{\vk'} \rangle = \mathrm{Tr}\, a^\dagger_{\vk} \, a_{\vk'}\,\rho(t_0) \equiv  n_k(0)\,\delta_{\vk,\vk'} = \frac{1}{e^{\beta_0 \hbar W_k}-1}\,\,\delta_{\vk,\vk'}~~;~~ \langle a_{\vk} a_{\vk'} \rangle =0  \,, \,\forall \vk,\vk' \,,\label{ocuzero}\eea and Wick's theorem applies.

\subsection{Equations of motion}\label{subsec:eom}

 After thermal decoupling, the density matrix is frozen in the Heisenberg picture and the time evolution is unitary, namely the dynamics must be studied as a \emph{closed quantum system } by  evolving the  density matrix (\ref{rhozeroin}) in time in the Schroedinger picture with the unitary time evolution operator.

 For any operator $\mathcal{O}$, Heisenberg's   equation of motion become
\be \frac{d}{dt}\mathcal{O}(\vx,t) = i \big[ H(t), \mathcal{O}(\vx,t)\big] \,,\label{Heiseom}\ee where we allow the Hamiltonian to depend explicitly on time. The  solution of (\ref{Heiseom}) is
\be \mathcal{O}(\vx,t_0) = U^{-1}(t,t_0)\,\mathcal{O}(\vx,t_0)\,U(t,t_0) \,,\label{solhem}\ee where the unitary time evolution operator (in what follows we set $\hbar=1$) is given by
\be U(t,t_0)= T\Big(e^{-i\int^t_{t_0}H(t')dt'}\Big)  ~~;~~ U^{-1}(t,t_0) = \widetilde{T}\Big(e^{i\int^t_{t_0}H(t')dt'}\Big)   \,.\label{Uop}\ee  where $T,\widetilde{T}$ are the time and anti-time ordering symbols.

 In the Heisenberg picture a density matrix does not depend on time, whereas in the Schroedinger picture its time evolution is given by
 \be \rho(t) = U(t,t_0)\,\rho(t_0)\, U^{-1}(t,t_0)\,,\label{rhot}\ee namely the density matrix evolves unitarily in time, as a consequence the entropy $ S = -\mathrm{Tr} \rho(t)\,\ln(\rho(t))$ is time independent.

 With an initial state described by a density matrix $\rho(t_0)$, normalized such that $\mathrm{Tr} \rho(t_0) =1$, expectation values of a Heisenberg field operator are given by
\be \langle \mathcal{O}(t) \rangle = \mathrm{Tr}\mathcal{O}(t) \,\rho(t_0) = \mathrm{Tr} \mathcal{O}(t_0)\,\rho(t)\,.\label{exval}\ee  Expectation values and correlation functions are obtained via functional derivatives of the generating functional\cite{beilok,boylee}
\be \mathcal{Z}[J^+,J^-] \equiv \mathrm{Tr}\Big[ U(t,t_0;J^+)\,\rho(t_0)\,U^{-1}(t,t_0;J^-)\Big] \,,\label{ZJ2}\ee with respect to the external sources $J^\pm$, where
\be   U(t,t_0;J^+) = \mathrm{T} \Big( e^{-i \int^t_{t_0}H(t';J^+)} \Big)~~;~~  U^{-1}(t,t_0;J^-)= \widetilde{T}\Big( e^{i\int^t_{t_0}\,H(t^{'};J^-)\,dt^{'}} \Big)\ee  with
\be H(t,J^{\pm}) \equiv H(t) + \int d^3 x J^{\pm}(\vx,t)\,\mathcal{O}(\vx,t) \,. \label{Hjs}\ee   For example correlation functions
\bea \langle   \mathcal{O}^+(\vx_1,t_1)\mathcal{O}^+(\vx_2,t_2) \rangle  &  =  & \mathrm{Tr}  \Big(T \mathcal{O}(\vx_1,t_1)\mathcal{O}(\vx_2,t_2)\Big) \rho(t_0) \nonumber \\ & = &  -\frac{\delta^2\,\mathcal{Z}[J^+,J^-]}{ \delta J^+(\vx_1,t_1)\delta J^+(\vx_2,t_2) } \Big|_{J^+=J^-=0}\,,  \label{timor} \eea
\bea
\langle \mathcal{O}^-(\vx_2,t_2)\mathcal{O}^+(\vx_1,t_1)\rangle  & = & \mathrm{Tr}      \mathcal{O}(\vx_1,t_1)\, \rho(t_0)\,\mathcal{O}(\vx_2,t_2) \nonumber \\ & = &  \frac{\delta^2\,\mathcal{Z}[J^+,J^-]}{ \delta J^+(\vx_1,t_1)\delta J^-(\vx_2,t_2) } \Big|_{J^+=J^-=0}\,, \label{timo}\eea etc. An important result is that
\bea  && \langle  \mathcal{O}^+(\vx,t)\rangle   \equiv     \mathrm{Tr}      \mathcal{O}(\vx,t)\, \rho(t_0)  = i\frac{\delta\,\mathcal{Z}[J^+,J^-]}{ \delta J^+(\vx,t)} \Big|_{J^+=J^-=0} \nonumber \\ & = &  \langle  \mathcal{O}^-(\vx,t)\rangle   \equiv    \mathrm{Tr}       \rho(t_0) \,  \mathcal{O}(\vx,t)= -i\frac{\delta\,\mathcal{Z}[J^+,J^-]}{ \delta J^-(\vx,t)}\Big|_{J^+=J^-=0} \,.\label{exvalj}\eea  This is the Schwinger-Keldysh or in-in formulation of non-equilibrium quantum field theory\cite{schwinger,keldysh,maha,jordan,beilok,boylee}.

Let us consider a scalar quantum field theory for a field $\phi$ as discussed in the previous sections,  the generating functional (\ref{ZJ2}) in the field representation can be written in a functional integral representation
\be \mathcal{Z}[J^+,J^-] = \int D\phi_f D\phi_i D\phi'_i \,\bra{\phi_f} U(t,t_0;J^+) \ket{\phi_i}\,\bra{\phi_i}\rho(t_0)\ket{\phi'_i}\bra{\phi'_i}U^{-1}(t,t_0;J^-)\ket{\phi_f}\,,\label{parti} \ee in turn the field matrix elements of the evolution operators can be written as path integrals, namely
\bea  \bra{\phi_f} U(t,t_0;J^+) \ket{\phi_i} & \equiv &  \int \mathcal{D}\phi^+ \,e^{i\int  \mathcal{L}[\phi^+;J^+]\,  d^4x} ~~;~~ \phi^+(t_0) = \phi_i;\phi^+(t)=\phi_f \,,\label{lplus} \\ \bra{\phi'_i} U^{-1}(t,t_0;J^-) \ket{\phi_f} & \equiv &  \int \mathcal{D}\phi^- \,e^{-i\int \mathcal{L}[\phi^-;J^-]\,  d^4x} ~~;~~ \phi^-(t_0) = \phi'_i;\phi^-(t)=\phi_f \,,\label{lmin}\eea where
\be \mathcal{L}[\phi^\pm;J^\pm] =  \tfrac{1}{2}  \,\Big(\frac{\partial \phi^\pm }{\partial t} \Big)^2 -\tfrac{1}{2} \,\bigl( \nabla \phi^\pm  \bigr)^2 - V(\phi^{\pm})-J^{\pm} \,\phi^{\pm} \,  \,.\label{lagspm}  \ee  Finally, the functional and path integral representation of the generating functional becomes
\be \mathcal{Z}[J^+,J^-] =   \int D\phi_f D\phi_i D\phi'_i \int \mathcal{D}\phi^+ \mathcal{D}\phi^- e^{i \int \Big[\mathcal{L}[\phi^+;J^+]-\mathcal{L}[\phi^-;J^-]\Big] d^4x} \,\rho(\phi_i,\phi'_i;t_0) \,,\label{pathint} \ee with the boundary conditions on the fields $\phi^\pm$ given by eqns. (\ref{lplus},\ref{lmin}) and the notation $\int d^4x \equiv \int^t_{t_0} dt' \int d^3 x$. The doubling of fields with the $\pm$ branches is a direct consequence of the time evolution of a \emph{density matrix}, with time evolution forward via $U(t,t_0)$ and backwards with $U^{-1}(t,t_0)$, in contrast to the usual S-matrix or in-out formulation which involves only time evolution forward because it evolves a state rather than a
density matrix.

Our objective is to obtain the equation of motion for the   expectation value of the   scalar field $\phi$, namely
\be \mathrm{Tr} \, \phi(\vx,t) \rho(t_0) \equiv \X(t) \,,\label{Xv}\ee  where we consider $\X$ to be spatially homogeneous,  hence only the zero momentum component of $\phi$ acquires an expectation value. The equation of motion for $\X$ is obtained by following the identity (\ref{exvalj}) which implies that $\langle \phi^+ \rangle = \langle \phi^- \rangle = \X$.

The equation of motion for $\varphi(t)$ is obtained by writing
\be \phi^\pm_1(\vx,t) = \X(t) + \delta^\pm(\vx,t)\,, \label{shift}\ee    in the Lagrangian $\mathcal{L}[\phi^\pm,J^\pm]$ in equation (\ref{lagspm}) and requesting that
\be \langle \delta^\pm(\vx,t)\rangle =0\,,\label{zerodelta}\ee  to all orders in perturbation theory, namely the same constraint as in the static case (\ref{constraint}).

Upon integration by parts and neglecting surface terms which do not contribute to equations of motion, and coupling sources only to the fluctuating fields $ \delta^\pm$,  we obtain (dots denote $\partial/\partial t$)
\bea && i\int  \Big[\mathcal{L}[\X;\delta^+, J^+] - \mathcal{L}[\X;\delta^-, J^-]\Big]  d^4x = \nonumber \\ &&  + i \int \frac{1}{2}\,\Bigg\{\Big(\frac{\partial \delta^+}{\partial t}\Big)^2-\Big( \nabla \delta^+ \Big)^2-V^{''}(\varphi(t))\,{\delta^+}^2+ J^+ \delta^+ \Bigg\} \nonumber \\ && -i \int \frac{1}{2}\,\Bigg\{\Big(\frac{\partial \delta^-}{\partial t}\Big)^2-\Big( \nabla \delta^- \Big)^2-V^{''}(\varphi(t))\,{\delta^-}^2+ J^-  \delta^-  \Bigg\}\,d^4 x \nonumber \\ && -i\int  \Bigg\{\Big(\ddot{\X}(t)+V^{'}(\varphi(t))\Big)\delta^+(\vx,t)+\frac{1}{3\!}\, {V^{'''}(\varphi(t))}  {\delta^+}^3 + \cdots \Bigg\}\,d^4 x -  \Big( \delta^+ \rightarrow \delta^- \Big)\nonumber \\
 \,.\label{shiftedlag}
 \eea

 The currents $J^\pm$ in this expression are intended to yield the correlation functions of the fluctuations $\delta^\pm$ in terms of functional derivatives with respect to them, and should not be confused with the Lagrange multiplier $j$ in the static case of the previous section which enforces the constraint (\ref{constraint}).

\begin{figure}[ht!]
\begin{center}
\includegraphics[height=3in,width=4in,keepaspectratio=true]{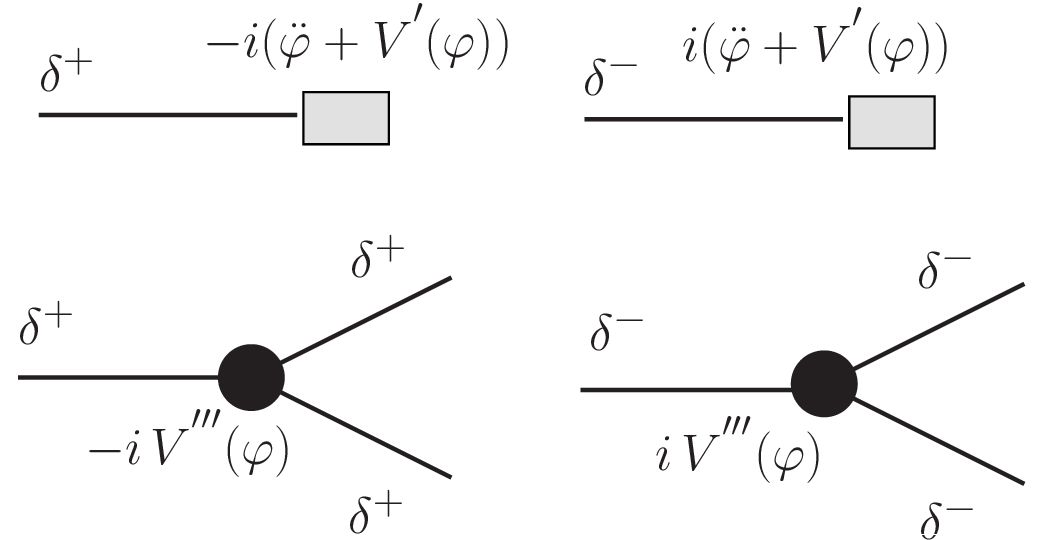}
\caption{Interaction vertices from the Lagrangian (\ref{shiftedlag}) up to and including $\mathcal{O}(\delta^3)$, the solid lines correspond to the fluctuations $\delta^\pm$. The gray box stands for $\mp i \Big(\ddot{\varphi}(t)+V^{'}(\varphi(t)) \Big)$, the black dot stands for $\mp i V^{'''}(\varphi(t))$.}
\label{fig:vertices}
\end{center}
\end{figure}

 The last line in (\ref{shiftedlag}) determines the interaction vertices, these are depicted in fig. (\ref{fig:vertices}),  just as in the static case,  the linear term is considered as part of the interaction. It is instructive to compare to the static case in particular the interaction term in eqn. (\ref{HI}), which shows that in the dynamical case $\ddot{\varphi}$ in the linear term in $\delta(\vx,t)$ in the interaction term in the last line in (\ref{shiftedlag}) replaces the Lagrange multiplier $J$ in (\ref{HI}). This is in agreement with the discussion right before the classical equation of motion  (\ref{eomclas}) comparing it to the constraint equation (\ref{extreV}).

The equation of motion for the mean field is obtained from the condition $\langle \delta^{\pm}(\vec{0},0) \rangle =0$. Considering the interaction term in (\ref{shiftedlag}) to first order, we find
\be -i \int \Bigg\{ \Big(\ddot{\X}(t)+V^{'}(\varphi)\Big) \,\langle \delta^+(\vec{0},0)\delta^+(\vx,t) \rangle + \frac{V^{'''}(\varphi(t))}{3\!}  \langle \delta^+(\vec{0},0)\,\big({\delta^+}(\vx,t)\big)^3 \rangle \Bigg\} d^4 x =0 \,. \label{eomdyn}\ee The expectation values are obtained in the free field theory defined by the first two lines in (\ref{shiftedlag}), with the initial density matrix $\rho(t_0)$.

The first two lines in (\ref{shiftedlag}) describe a free scalar field theory with a time dependent mass $V^{''}(\varphi(t))$, yielding the field equations (\ref{heis}), and the field expansion (\ref{delex}).

 Using Wick's theorem it follows that
 \be \langle \delta^+(\vec{0},0)\,\big({\delta^+}(\vx,t)\big)^3 \rangle = 3 \, \langle \delta^+(\vec{0},0)\, \delta^+ (\vx,t)  \rangle\, \langle \big({\delta^+}(\vx,t)\big)^2 \rangle \,, \label{wicki}\ee
 and factorizing $\langle \delta^+(\vec{0},0)\delta^+(\vx,t) \rangle $  from the expression  (\ref{eomdyn}) we find
\be   \ddot{\X}(t)+V^{'}(\varphi)  \, + \frac{1}{2}  V^{'''}(\varphi(t))   \langle \Big( {\delta^+}(\vx,t)\Big)^2  \rangle = 0 \,,\label{eomfindyn} \ee   this equation is depicted symbolically in fig. (\ref{fig:eom}).

Various correlation functions needed to obtain the equations of motion are summarized in appendix (\ref{app:corre}).

 \begin{figure}[ht!]
\begin{center}
\includegraphics[height=3in,width=4in,keepaspectratio=true]{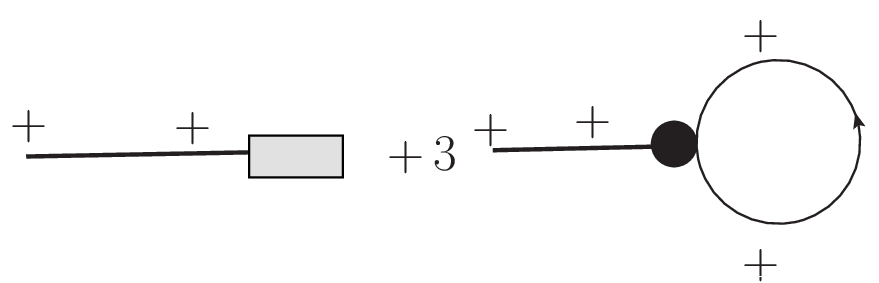}
\caption{Equation of motion up to one loop. The solid straight line is the propagator $\langle \delta^+(\vec{0},0)\delta^+(\vx,t) \rangle $, the closed loop is
$\langle \delta^+(\vec{x},t)\delta^+(\vx,t) \rangle $.}
\label{fig:eom}
\end{center}
\end{figure}

With the field expansion (\ref{delex}) and the correlations (\ref{ocuzero}) it is straightforward to find
\be  \langle \Big( {\delta^+}(\vx,t)\Big)^2  \rangle = \hbar\,\int \frac{d^3k}{(2\pi)^3} |g_k(t)|^2\, \Big(1+ 2 n_k(0)\Big) \,,\label{hart}\ee finally yielding the equation of motion
\be \ddot{\X}(t)+V^{'}(\varphi(t))  \, +  \frac{ \hbar }{2}\,  V^{'''}(\varphi(t))\, \int \frac{d^3k}{(2\pi)^3} |g_k(t)|^2\, \Big(1+2 n_k(0) \Big)=0 \,.\label{fineomdyn}  \ee

It is straightforward to check that the same equation of motion is obtained from using the backward branch ($\delta^-$) contribution from the last line in eqn. (\ref{shiftedlag}). This is because $\langle \big({\delta^+}(\vx,t)\big)^2 \rangle = \langle \big({\delta^-}(\vx,t)\big)^2 \rangle$ where now the propagator $\langle \delta^+(\vec{0},0)\delta^-(\vx,t) \rangle $ is factorized. Also, and of course, the same equation is obtained by considering the constraint $\langle \delta^-(\vec{0},0) \rangle =0$.

This method to obtain the equations of motion for expectation values, based on the in-in or Schwinger-Keldysh formulation of non-equilibrium quantum field theory is general and applies to any quantum field theory, furthermore, with few modifications it can be extended to the realm of cosmology\cite{cosmocao}.

In the case of the scalar field theory defined by the Hamiltonian (II.1) the equation of motion (IV.38) can also be obtained directly from the Heisenberg field equation, which follow from the variational principle applied to the full action

\be \ddot{\phi}(\vx,t)-\nabla^2\phi(\vx,t) + V^{'}(\phi(\vx,t)) =0 \,,\label{heisi} \ee which is obviously  fulfilled as an expectation value in the initial density matrix, namely
\be \mathrm{Tr} \rho(t_0) \Bigg( \ddot{\phi}(\vx,t)-\nabla^2\phi(\vx,t) + V^{'}(\phi(\vx,t))     \Bigg) =0 \,.\label{iniexpval}\ee

Shifting the field operator by the spatially homogeneous mean field $\phi(\vx,t) = \varphi(t)+\delta(\vx,t)$
yields, for the Heisenberg field equation (\ref{heisi})
\be \ddot{\varphi}(t)+ V^{'}(\varphi(t))+\Big[ \ddot{\delta}(\vx,t)-\nabla^2\delta(\vx,t)+V^{''}(\varphi(t))\delta(\vx,t)\Big]+ \frac{1}{2}V^{'''}(\varphi(t))\,\delta^2(\vx,t) + \cdots =0 \,,\label{heisidel}\ee   using the quantization of the fluctuation via the solution of the free field equations of motion in the background of the mean field, equations (\ref{heis},\ref{delex}),    leads to the vanishing of the (third) term inside the bracket in (\ref{heisidel}), yielding the expectation value (\ref{iniexpval})
\be  \ddot{\varphi}(t)+ V^{'}(\varphi(t))+ \frac{1}{2}V^{'''}(\varphi(t))\,\Big(\mathrm{Tr}\rho(t_0) \delta^2(\vx,t)\Big) + \cdots =0 \,.\label{expheisi}\ee With the initial density matrix given by (\ref{rhozeroin}), the field expansion (\ref{delex}) and the expectation values (\ref{ocuzero}),  it is straightforward to find that
\be \mathrm{Tr}\rho(t_0) \delta^2(\vx,t) = \hbar \int \frac{d^3k}{(2\pi)^3} |g_k(t)|^2\, \Big(1+2 n_k(0) \Big)\,.\label{trasi} \ee

Thereby confirming the equation of motion (IV.38) obtain via the more general in-in Schwinger-Keldysh formulation. This is not only reassuring, but also confirms that the equations of motion for the condensate follow from unitary time evolution as it is obtained from the expectation value of the Heisenberg equations of motion for the field operators in a time independent density matrix. While this latter derivation is arguably simpler, we have also presented the more rigorous in-in formulation due to its generality and appropriate use in non-equilibrium quantum field theory.

In references\cite{beilok,brownian,axionbrownian} the Schwinger-Keldysh effective action up to one loop was obtained in terms of the fields with labels $\pm$ on the two branches, corresponding to forward and backwards time evolution. This action is re-written in terms of the Keldysh center of mass and relative variables $(\Phi =\phi^++\phi^-)/2;R=(\phi^+-\phi^-)$ respectively and   the equations of motion obtained from a variational principle on these variables\cite{beilok}. Up to one-loop it is shown in refs.\cite{brownian,axionbrownian} that the effective equations of motion are of the Langevin type, with a Gaussian stochastic noise, the expectation value of the scalar field is directly determined by the expectation value of the center of mass coordinate with the probability distribution function (Gaussian) of this noise, yielding the equations of motion for the expectation value of the scalar field in the initial density matrix\cite{brownian,axionbrownian}, which is the method used above to derive the equations of motion from the Heisenberg field equations.

When the mean field $\varphi$ is time independent, namely in the static case, and when $\varphi$ is away from the spinodal region the mode functions are
\be g_k(t) = \frac{e^{-i\omega_k t}}{\sqrt{2\omega_k}} ~~;~~ \omega_k = \sqrt{k^2+V^{''}(\varphi)}\,,\label{gkficons2}\ee (see eqn. (\ref{gkficons})) and the last two terms in the equation of motion (\ref{fineomdyn}) become
\be V^{'}(\varphi(t))  \, +  \frac{ \hbar }{2}\,  V^{'''}(\varphi(t))\, \int \frac{d^3k}{(2\pi)^3} \frac{1}{2\omega_k}\, \Big(1+2 n_k(\varphi) \Big) = \frac{d}{d\varphi}V_{eff}(\varphi)\,, \ee in agreement with the static case, eqn. (\ref{jotah}). However, when $\varphi(t)$ is dynamical, the mode functions $g_k(t)$ describe the parametric and spinodal instabilities discussed in the previous sections and the last two terms in the equation of motion (\ref{fineomdyn}) cannot be identified with a derivative of an effective potential.

As a consequence of  the mode equations (\ref{modeg}), it is straightforward to show that the equation of motion (\ref{fineomdyn})  yields the conserved quantity
\be \widetilde{\mathcal{E}} = \underbrace{\frac{1}{2}\,(\dot{\varphi}(t))^2 + V(\varphi(t)) }_{ \widetilde{\mathcal{E}}_{cl} }+ \underbrace{\frac{ \hbar }{2} \, \int \frac{d^3k}{(2\pi)^3}\Big[|\dot{g}_k(t)|^2+\omega^2_k(t)\,|g_k(t)|^2\Big]\, \Big(1+2 n_k(0) \Big)}_{\widetilde{\mathcal{E}}_{fl}}= \mathrm{constant}\,,\label{consE}\ee as can be easily confirmed by taking $\dot{\widetilde{\mathcal{E}}}$ and    using   equations (\ref{modeg}),  yielding $\dot{\varphi}$ times equation   (\ref{fineomdyn}). The brackets in eqn. (\ref{consE}) define the  classical ($\widetilde{\mathcal{E}}_{cl}$), and fluctuation ($\widetilde{\mathcal{E}}_{fl}$) contributions to the total energy density respectively.

It is important, and enlightening, to compare this conservation law to that obtained from using the \emph{static} effective potential in the dynamical equation of motion of the homogeneous mean field (\ref{Econs}).  First, we  prove that the equation (\ref{consE}) is the expectation value of the \emph{time independent} Hamiltonian (\ref{Hamiltonian1}) in the initial density matrix $\rho(t_0)$ up to $\mathcal{O}(\hbar)$, namely one loop order. Let us shift both the field $\phi$ and its canonical momentum $\pi$ as
\be \phi(\vx,t) = \varphi(t)+\delta(\vx,t)~~;~~ \pi(\vx,t) = \dot{\varphi}(t)+\pi_{\delta}(\vx,t) \,,\label{shiftfipi}\ee yielding
\be H[\phi] = \mathcal{V}\Big[ \frac{1}{2} (\dot{\varphi}(t))^2 + V(\varphi(t))\Big]+ H_{\delta} + \cdots \,,\label{hspli}\ee where
\be H_{\delta} = \int d^3x \Bigg\{\frac{\hat{\pi}^2_{\delta}}{2}+\frac{(\nabla \hat{\delta})^2}{2} + \frac{V''(\varphi(t))}{2}\,\hat{\delta}^2 \Bigg\}\,, \label{hdeli}\ee   and the dots in equation (\ref{hspli}) stand for linear terms with vanishing expectation value in the density matrix $\rho(t_0)$, along with cubic and quartic terms in $\delta$ which yield higher loop corrections. Upon quantization of the fluctuation field via the mode expansion (\ref{delex},\ref{pidelex})  and using  the expectation values  (\ref{ocuzero}), we  find
  \be \frac{1}{\mathcal{V}} \,\mathrm{Tr}H_{\delta}\,\rho(t_0) = \frac{\hbar}{2}\int \frac{d^3k}{(2\pi)^3}\,\Big[ |\dot{g}_k(t)|^2 + \omega^2_k(t)\,|g_k(t)|^2 \Big]\,\Big(1+2 n_k(0) \Big)  \,,\label{exphdeli}\ee therefore, up to one-loop ($\mathcal{O}(\hbar)$) we find
  \be \widetilde{\mathcal{E}} = \frac{1}{\mathcal{V}} \,\mathrm{Tr}H[\phi]\,\rho(t_0) \,,\label{equival}\ee namely the constancy of $\widetilde{\mathcal{E}}$ is the statement that the field Hamiltonian is time independent. In contrast to equation (\ref{Econs}) with the caveats discussed in the previous section in the broken symmetry case, the expectation value of the energy density is constant and \emph{always real}, and because the time evolution is unitary the entropy density
   \be \mathcal{S} = \int \frac{d^3 k}{(2\pi)^3}\,\Big\{(1+n_k(0))\,\ln (1+n_k(0))- n_k(0)\,\ln (n_k(0))\Big\}\,, \label{Seqfroz} \ee  where equation (\ref{Seqfroz}) is obtained using the initial thermal state set by the decoupling and  is   constant and real.

   This is in striking contrast to equation (\ref{Econs}) with $V_{eff}(\varphi)$ given by equation (\ref{Sdef}) in terms of the internal energy and entropy densities (\ref{Uen},\ref{Seq}) each one varying in time, with a non-monotonic behavior for the entropy and both featuring an imaginary part when the mean field is in the spinodal region.

\subsection{Stimulated particle production:}\label{subsec:stimpp}

The equation of motion (\ref{fineomdyn}) and conservation law (\ref{consE}) are very similar to the zero temperature case obtained in ref.\cite{veff}, with the only difference being the initial  occupation number in the one-loop contribution. Following this reference, this similarity suggests us to relate the growth of the mode functions either by parametric amplification or spinodal instabilies to particle production.

\textbf{Unbroken symmetry case:} in this case the time dependent frequencies $\omega_k(t) = \sqrt{k^2+V^{''}(\varphi(t))}$ are always positive, and we introduce the zeroth adiabatic order mode functions
\be \tf_k(t)= \frac{e^{-i\,\int^{t}\,\omega_k(t')\,dt'}}{\sqrt{2\,\omega_k(t)}} \,.  \label{zerof}\ee We expand the \emph{exact} mode functions $g_k(t)$ in terms of these adiabatic modes by introducing Bogoliubov coefficient functions $\ta_k(t),\tb_k(t)$ defined by the following relations
\bea g_k(t) & = & \ta_k(t)\,\tf_k(t)+ \tb_k(t)\,\tf^*_k(t) \label{gexpa}\\
  \dot{g}_k(t) & = &  -i \omega_k(t)\,\Big[  \ta_k(t)\,\tf_k(t)-\tb_k(t)\,\tf^*_k(t)\Big] \,. \label{dergexpa}\eea
which  can be inverted to obtain the Bogoliubov coefficients
\bea \ta_k(t) & = &  i\,\tf^*_k(t) \Big[\dot{g}_k(t) -i \omega_k(t)  \,g_k(t)  \Big] \label{tilA} \\
\tb_k(t) & = &  -i\,\tf_k(t) \Big[\dot{g}_k(t) +i \omega_k(t) \,g_k(t)  \Big]\,. \label{tilB}\eea

It follows from  the Wronskian condition (\ref{wron}) that
\be |\ta_k(t)|^2-|\tb_k(t)|^2 =1\,. \label{Wroab}\ee

 The definitions (\ref{gexpa}, \ref{dergexpa})  yield
\bea a_{\vk}\,g_k(t) + a^\dagger_{-\vk} \, g^*_k (t)  & = & c_{\vk}(t) \,\tf_k(t)+  c^{\dagger}_{-\vk}(t)\,\tf^*_k(t)\,, \label{adiaexp}\\
a_{\vk}\,\dot{g}_k(t) + a^\dagger_{-\vk} \, \dot{g}^*_k (t)  & = & -i\omega_k(t) \,\Big( c_{\vk}(t) \,\tf_k(t)-  c^{\dagger}_{-\vk}(t)\,\tf^*_k(t)\Big)\,, \label{dotadiaexp}\eea where
\be c_{\vk}(t) = a_{\vk}\,\ta_k(t)+a^\dagger_{-\vk}\,\tb^*_k(t)~~;~~c^\dagger_{\vk}(t) = a^\dagger_{\vk}\,\ta^*_k(t)+a_{-\vk}\,\tb_k(t)\,, \label{cops} \ee the condition (\ref{Wroab}) ensures that $c_{\vk}(t);c^\dagger_{\vk}(t)$ obey equal time canonical commutation relations. It is straightforward to show that the quadratic Hamiltonian $H_\delta$ given by eqn. (\ref{hdeli}) can be written in terms of the time dependent operators $c^{\dagger}_{\vk}(t);c_{\vk}(t)$ as
\be H_{\delta} = \sum_{\vk} \hbar \omega_k(t) \Big[ c^{\dagger}_{\vk}(t)\,c_{\vk}(t)+\frac{1}{2}\Big]\,.\label{adiH}\ee Following ref.\cite{veff} we \emph{define} the number of adiabatic particles as
\be \widetilde{\mathcal{N}}_k (t) = \bra{0}c^\dagger_{\vk}(t)\, c_{\vk}(t)\ket{0} = |\tb_k(t)|^2 \,, \label{adianum}\ee where the \emph{vacuum} state $\ket{0}$ is such that
\be a_{\vk}\ket{0} =0 ~~;~~\forall{\vk} \,.\label{vcu}\ee
The relation (\ref{tilB}) and the Wronskian condition (\ref{wron})   yield
  \be \widetilde{\mathcal{N}}_k (t) = \frac{1}{2\omega_k(t)}\Big[|\dot{g}_k(t)|^2+ \omega^2_k(t) |g_k(t)|^2\Big]- \frac{1}{2}\,, \label{Ngrel}\ee from which it follows that
\be \frac{1}{\mathcal{V}} \,\mathrm{Tr}H_{\delta}\,\rho(t_0) = \frac{\hbar}{2} \int \frac{d^3k}{(2\pi)^3}\,\omega_k(t)\Big[1+2\widetilde{\mathcal{N}}_k (t)\Big] \Big(1+2 n_k(0) \Big) \,.\label{exvalHdel}\ee

With the initial conditions (\ref{inigis}) ($g_k(0)= \frac{1}{\sqrt{2\omega_k(0)}}~;~ \dot{g}_k(0) = -i\frac{\omega_k(0)}{\sqrt{2\omega_k(0)}}$), it follows that
\be \widetilde{\mathcal{N}}_k (0) = 0\,,\label{nopart}\ee therefore the initial state is the vacuum state for the adiabatic  particles. The \emph{distribution function} for the adiabatic particles is given by
\be \mathcal{F}_k(t) = \mathrm{Tr} \Big(c^\dagger_{\vk}(t) c_{\vk}(t) \,\rho(t_0) \Big)= \widetilde{\mathcal{N}}_k (t)+ n_k(0)\Big[1+2\widetilde{\mathcal{N}}_k (t)\Big] ~~;~~ \mathcal{F}_k(0)= n_k(0)\,,\label{stimupp}\ee the second term in $\mathcal{F}_k(t)$ describes \emph{stimulated}   production of adiabatic particles. In terms of this distribution function, the one loop contribution to the energy density, eqn. (\ref{exvalHdel}) can be written in the following illuminating manner,
\be  \frac{1}{\mathcal{V}} \,\mathrm{Tr}H_{\delta}\,\rho(t_0) =  \frac{\hbar}{2} \int \frac{d^3k}{(2\pi)^3}\,\omega_k(t)\Big[1+2\mathcal{F}_k(t)\Big]\,.  \label{enepro}\ee We can now gather these results to express the conserved energy density (\ref{consE}) in the form
\be \widetilde{\mathcal{E}} = \frac{1}{2}\,(\dot{\varphi}(t))^2 + V(\varphi(t)) + \frac{ \hbar }{2} \, \int \frac{d^3k}{(2\pi)^3}\omega_k(t)\Big[1+2\mathcal{F}_k(t)\Big] \,.\label{consEpp}\ee This expression is remarkably similar to the energy density obtained at zero temperature in ref.\cite{veff}, but in terms of the distribution function $\mathcal{F}_k(t)$, which describes stimulated particle production instead of the vacuum adiabatic particle number density $\widetilde{\mathcal{N}}_k (t)$. The conservation of $\widetilde{\mathcal{E}}$   along with eqn. (\ref{stimupp}) taken together have an important physical interpretation of the dynamics:  a mechanism of \emph{energy transfer} between the mean field and the quantum fluctuations resulting in the stimulated production of the adiabatic particles with \emph{non-thermal} distributions. In particular, the exponential growth of the mode functions $g_k(t)$ as a consequence of parametric amplification must result in a drain of the energy stored in the mean field, energy that goes into particle production with non-thermal distributions.

The motivation for the choice of the zeroth-order adiabatic mode functions (\ref{zerof})  now becomes clear: while $\varphi(t)$ is oscillating around the minimum, parametric amplification of fluctuations drains energy from the condensate, diminishing its amplitude. This dissipative mechanism entails that asymptotically $\varphi$ will settle at the minimum and the frequencies become slowly varying functions of tima approaching an asymptotic limit $\omega_k(\infty)=\sqrt{k^2+V^{''}(\varphi=0)}$. In this limit the mode functions $\tf_k(t)\rightarrow   {e^{-i\,\omega_k(\infty) t}}/{\sqrt{2\,\omega_k(\infty)}}  $ describing asymptotic ``out'' particle states.

\vspace{2mm}

\textbf{Broken symmetry case:} This case is more subtle. Although it is not clear that the fluctuation contribution $\widetilde{\mathcal{E}}_{fl}$ in equation (\ref{consE}) grows as a consequence of the spinodal instabilities, since for spinodally unstable modes $\omega^2_k(t) < 0$, it follows from the mode equations (\ref{modeg}) that
\be \dot{\widetilde{\mathcal{E}}_{fl}} = \frac{\hbar}{2}\,\frac{d}{dt}\Big(V^{''}(\varphi(t))\Big)\,\int \frac{d^3k}{(2\pi)^3}\,|g_k(t)|^2\,\Big(1+2 n_k(0) \Big) \,,\label{dotefl}\ee As $\varphi(t)$ rolls down the potential hill from near the maximum of the potential towards the symmetry breaking minima, $V^{''}(\varphi(t))$ \emph{increases} from a negative value to zero at the inflection point, namely the end of the spinodal region. Therefore, because $|g_k(t)|^2$ grows nearly exponentially in this region, it follows that the fluctuation contribution grows nearly exponentially while $\varphi(t)$ traverses the spinodal region. Furthermore, the temperature correction in (\ref{dotefl}) implies an enhancement as compared to the zero temperature case\cite{veff}, again a manifestation of stimulated  production of fluctuations. Because the total energy density remains constant, this energy is drained from the classical contribution $\widetilde{\mathcal{E}}_{cl}$ in equation (\ref{consE}), again, a mechanism of energy transfer from the mean field to the fluctuations implying damping of the amplitude of the mean field. Because $\omega_k(t)$ are imaginary for spinodally unstable wavevectors, we cannot define the adiabatic modes as in the previous case. However, motivated by the argument that the growth of fluctuations implies a damping of the mean field as a consequence of energy transfer to the flutuations, we follow the treatment of ref.\cite{veff}  and introduce $K_s$ as the maximum unstable wavevector  while $\varphi(t)$ is in the spinodal region. For example for the   typical potential $V(\varphi) = -m^2\,\varphi^2/2 + \lambda \varphi^4/4$ witn $m^2 >0$, it follows that the maximum unstable wavevector is
\be K_s = |V^{''}(0)|\,.\label{Ks}\ee
For $k\leq K_s$ there is no unambiguous definition of an adiabatic particle number, whereas for $k > K_s$ the mode functions can again be written as in eqn. (\ref{gexpa},\ref{dergexpa}) in terms of the zeroth order adiabatic modes yielding the results obtained above for the case of unbroken symmetry. Therefore, separating the spinodally unstable modes we now write the fluctuation contribution to the energy density (\ref{consE}) as
\bea \widetilde{\mathcal{E}}_{fl} & = &  \frac{\hbar}{4\pi^2} \, \int^{K_s}_0  \Big[|\dot{g}_k(t)|^2+\omega^2_k(t)\,|g_k(t)|^2\Big]\, \Big(1+2 n_k(0) \Big)\,k^2\,dk \nonumber \\
& + &  \frac{\hbar}{4\pi^2} \,\int^{\Lambda}_{K_s} \omega_k(t)\Big[1+2\mathcal{F}_k(t)\Big]\,k^2\,dk    \,.\label{flucE2}  \eea  where the distribution function $\mathcal{F}_k(t)$ is the same as in equation (\ref{stimupp}), and we have introduced an upper momentum cutoff $\Lambda \gg |V^{''}(\varphi)|$ to discuss renormalization aspects.

Since both spinodal and parametric instabilities lead to an efficient transfer of energy from the condensate to the fluctuations, we expect that at long time, the condensate will oscillate around a minimum below the classical spinodal as the instabilities eventually must shut off by energy conservation. In this asymptotic long time limit the $V^{''}(\varphi(t)) >0$ and the frequencies are real, and the contribution from the modes with $k< K_s$ becomes of the same form as for those with $k>K_s$. Therefore we expect that in the long time limit as $\varphi$ oscillates with  small amplitude  around a minimum away from and not probing the spinodal region, the mode functions can again be written as in equations (\ref{gexpa},\ref{dergexpa}), with the interpretation of asymptotic adiabatic particle production, so that the exponential growth from spinodal instabilities is imprinted in the Bogoliubov coefficient functions, thereby describing the production of asymptotic particles.

 Therefore, in this limit  both contributions in (\ref{flucE2}) have the same form in terms of the stimulated distribution function of produced particles $\mathcal{F}_k(t)$.

\subsection{Renormalized dynamical framework:}\label{subsec:renor}
The expression (\ref{flucE2}) for $\widetilde{\mathcal{E}}_{fl}$ allows us to treat both cases with and without symmetry breaking on the same footing: the case $K_s=|V^{''}(0)|\neq 0$ corresponds to symmetry breaking and $K_s =0$ to unbroken symmetry.  In ref.\cite{veff} the renormalization aspects were studied for the zero temperature case, which can be obtained from the results above by setting $n_k(0) =0$. Because of the exponential suppression of the high momentum modes in the thermal distribution functions, it follows that the ultraviolet divergences are those of the zero temperature case and, as discussed in detail in ref.\cite{veff},  are completely described by the ``1'' in the bracket in the second term in (\ref{flucE2}), namely the zero point energy.

We proceed to subtract this term from the fluctuation energy and lump it together with $V(\phi(t))$ in the full energy density (IV.44), thus defining a \emph{new effective potential}
\be \overline{ V}_{eff}(\varphi)   \equiv    V(\varphi) + \frac{\hbar}{4\pi^2} \,\int^{\Lambda}_{K_s} \omega_k(t)\,k^2 dk\,,\label{overV}\ee yielding
\bea \overline{ V}_{eff}(\varphi) & = &  V(\varphi) + \frac{\hbar}{16\pi^2} \left\{ {\Lambda^4}  + V^{''}(\varphi)\, {\Lambda^2} - \frac{1}{ 4}\,(V^{''}(\varphi))^2\,\Big[\ln\Big( \frac{4\Lambda^2}{\mu^2}\Big)-\frac{1}{2}\Big]+\frac{1}{ 4}\,(V^{''}(\varphi))^2 \, \ln\Big( \frac{|V^{''}(\varphi)|}{\mu^2}\Big)\right. \,\nonumber \\
 & - & \left. \Big(V^{''}(\varphi) \Big)^2\,\mathcal{H}\Bigg[\frac{K_s}{|V^{''}(\varphi) |^{1/2}} \Bigg] \right\}  \,,\label{vefina} \eea  where $\mu^2$ is a renormalization scale and
 \bea \mathcal{H}\big[x\big] & = & \frac{1} {2} \Bigg\{2\,x\, \Big[x^2+ \mathrm{sign}\big(V^{''}(\varphi)\big)\Big]^{3/2}  - x\,\mathrm{sign}\big(V^{''}(\varphi)\big)\,\Big[x^2+ \mathrm{sign}\big(V^{''}(\varphi)\big)\Big]^{1/2}\nonumber \\ & - &\ln\Big[x+\Big[x^2+ \mathrm{sign}\big(V^{''}(\varphi)\big)\Big]^{1/2} \Big] \Bigg\}\,,\label{funx}\eea with $K_s=0$ for unbroken symmetry and $K_s=|V^{''}(0)|$ for broken symmetry. In a renormalizable theory, the ultraviolet divergent terms are absorbed into renormalization of the parameters, for example for the bare scalar potential
  \be V(\varphi) = V_0+\frac{m^2_0}{2}\,\varphi^2+ \frac{\lambda_0}{4}\,\varphi^4   \,, \label{treepot}\ee
  renormalization is achieved by introducing the renormalized parameters
   \bea \frac{m^2_R(\mu)}{2} & = &  {m^2_0} + \frac{6\lambda_0}{16\pi^2}\,\Lambda^2-\frac{6\lambda_0}{32\pi^2}\,m^2_0\,\Big[\ln\Big( \frac{4\Lambda^2}{\mu^2}\Big)-\frac{1}{2}\Big]\,\label{mren}\\
 {\lambda_R(\mu)}  & = &  {\lambda_0} - \frac{36\,\lambda^2_0}{32\pi^2}\, \Big[\ln\Big( \frac{4\Lambda^2}{\mu^2}\Big)-\frac{1}{2}\Big] \,,\label{lambdaren}\\
V_{0R}(\mu) & = & V_0 + \frac{\Lambda^4}{16\pi^2}+m^2_0\, \frac{\Lambda^2}{16\pi^2}-\frac{m^4_0}{64\pi^2}\,\Big[\ln\Big( \frac{4\Lambda^2}{\mu^2}\Big)-\frac{1}{2}\Big] \,, \label{voren} \eea  and replacing bare by renormalized quantities up to one loop,
\be \overline{ V}^R_{eff}(\varphi)   =   V_R(\varphi) + \frac{1}{ 4}\,(V^{''}_R(\varphi))^2 \, \ln\Big( \frac{|V^{''}_R(\varphi)|}{\mu^2}\Big)- \Big(V^{''}_R(\varphi) \Big)^2\,\mathcal{H}\Bigg[\frac{K_{s_R}}{|V^{''}_R(\varphi) |^{1/2}} \Bigg]\,,\label{vefre}\ee where the subscript $R$ refers to the renormalized quantities in terms of the renormalized mass and coupling. The renormalization group invariance of the effective potential has been discussed in refs.\cite{colewein,veff}. We note that the argument of the function $\mathcal{H}$, is $\Big(|V^{''}_R(0)|/|V^{''}_R(\varphi) |\Big)^{1/2} >1$ for the broken symmetry case when $\varphi$ is within the spinodal region ($\mathrm{sign}(V^{''}(\varphi))<0$).

This effective potential is manifestly real, unlike the usual effective potential that becomes complex when $\varphi$ is within the spinodal region. After renormalization the total conserved energy density becomes
 \bea \widetilde{\mathcal{E}} & = &  \frac{1}{2}\,(\dot{\varphi}(t))^2 +  \overline{ V}^R_{eff}(\varphi) +\frac{\hbar}{4\pi^2} \, \int^{K_{s}}_0  \Big[|\dot{g}_k(t)|^2+\omega^2_k(t)\,|g_k(t)|^2\Big]\, \Big(1+2 n_k(0) \Big)\,k^2\,dk \nonumber \\
 & + &    \frac{\hbar}{2\pi^2} \,\int^{\Lambda}_{K_{s}} \omega_k(t) \,\mathcal{F}_k(t) \,k^2\,dk  \,,\label{renconsEpp}\eea where everywhere the mass and coupling are the renormalized quantities. The fully renormalized equations of motion are obtained as follows: beginning with the conserved energy density (\ref{consE}),  and $\mathcal{E}_{fl}$ given by eqn. (\ref{flucE2}) subtract from this expression the term with the ``1'' inside the bracket of the second line, and  lump it together with $V(\varphi)$ to define  $\overline{ V}^R_{eff}(\varphi)$ as in eqn. (\ref{overV}). Now taking the time derivative of $\mathcal{E}$ yields $\dot{\varphi}$ times the equation of motion, which upon using the equations for the mode functions (\ref{modeg})
 lead to the renormalized equation of motion
  \be \ddot{\varphi}(t)+\frac{d}{d\varphi}\overline{ V}^R_{eff}(\varphi) \, +  \frac{ \hbar }{4\pi^2}\,  V^{'''}_R(\varphi(t))\, \int^{\Lambda}_0  \Bigg\{ |g_k(t)|^2\,
  \Big(1+2 n_k(0) \Big)-\frac{\Theta(k-K_s)}{2\omega_k(\varphi)}\Bigg\} k^2 \,dk  =0 \,,\label{fineomren}  \ee where again, everywhere, the mass and coupling are the renormalized ones. The equation (\ref{fineomren}) along with the mode equations (\ref{modeg}) with initial conditions (\ref{sumainiconds}) provide a complete  description of the dynamics of the mean field (condensate) with the following properties:

  \begin{itemize}
  \item{ The equation of motion (\ref{fineomren}) is consistently renormalized.}

  \item{The renormalized effective potential $\overline{V}_R(\varphi)$ is manifestly always  \emph{real} for all values of the mean field even within the spinodal region, unlike the usual effective potential which is complex in the case when the tree level potential features broken symmetry minima. }

   \item{The energy density is manifestly real and conserved.}

  \item{ The equation of motion for the condensate arises from unitary time evolution of an initial density matrix, as confirmed by obtaining it also from the expectation value of the equations of motion of the Heisenberg field operators in the initial density matrix. Therefore the \emph{thermodynamic entropy is constant}.\footnote{The thermodynamic entropy should \emph{not} be confused with the coarse-grained \emph{entanglement entropy} discussed in ref.\cite{veff}.} }

  \end{itemize}

\section{Discussion}\label{sec:discussion}
The   dynamics described by the equation of motion (\ref{fineomren}) with the conserved energy (\ref{renconsEpp}) suggests the emergence of  stationary asymptotic states. Let us consider first the case in which the tree level potential features only one minimum, namely unbroken symmetry, with large amplitude initial conditions on the condensate. As $\varphi$ oscillates around the minimum parametric instabilities lead to profuse particle production, which drains energy from the ``classical'' part of the energy into the fluctuations, populating parametrically unstable bands in momentum with a non-thermal distribution function. Particle production will continue as long as oscillations continue as demonstrated with the simple Mathieu equation analysis in the previous section. As the energy of the condensate is drained from particle production, the amplitude of oscillations diminishes and the bandwidths of the unstable bands become narrower, suggesting a dissipative mechanism that drives the condensate to the equilibrium minimum but with a highly excited non-thermal population of particles. Eventually this transfer of energy must stop and $\varphi$ settles at the minimum with vanishing velocity, the frequencies $\omega_k(t)\rightarrow \omega_k(\infty)$,  and the zeroth-order mode functions (\ref{zerof}) describe asymptotic ``out'' single particle states. This is an asymptotic fixed point of the dynamics.

Such asymptotic limit will yield the asymptotic value(s) $\varphi(\infty)$ as the solution(s) of the renormalized equation of motion (\ref{fineomren}), subject to the constraint of total energy density (\ref{renconsEpp}) with $\dot{\varphi}(\infty)=0,\ddot{\varphi}(\infty)=0$.

If the tree level potential features symmetry breaking minima and the initial value of the mean field is large, with a large energy density, both spinodal and parametric instabilities will be effective in draining energy from the condensate leading to particle production with non-thermal distributions. As the amplitude of the mean field diminishes the mean field can asymptotically settle in a broken symmetry minimum away from the origin, but it is also possible, with a large energy density, that asymptotically the mean field settles in a state with vanishing value. This would imply a restoration of symmetry, which is a possibility for a large energy density, that must be studied numerically and will likely depend on the particular value of parameters. However, in this case the condensate oscillates around a minimum with diminishing amplitude eventually settling at this minimum and again the frequencies $\omega_k(t) \rightarrow \omega_k(\infty)$ and the zeroth-adiabatic order mode functions (\ref{zerof}) describe asymptotic ``out'' single particle states. In this case the Bogoliubov coefficients and the stimulated distribution function  (\ref{stimupp}) include the growth of fluctuations from both, spinodal and parametric instabilities. The asymptotic value(s)   $\varphi(\infty)$ are again determined by the solutions of the equation of motion (\ref{fineomren}) with the energy constraint (\ref{renconsEpp}) with $\dot{\varphi}(\infty)=0,\ddot{\varphi}(\infty)=0$.

When the amplitude of oscillations diminishes from the energy transfer to fluctuations via particle production, it is possible that the dynamics ``un-freezes'' and the coupling to the heat bath or alternative collisional processes become effective again, perhaps leading to a re-distribution of the produced quanta and a ``re-thermalization'' on longer time scales. At this stage, this is, of course, a conjecture that can only be
assessed with a detailed treatment of the quantum kinetics including the couplings to the bath and or other collisional processes, and merits further and deeper study.

\section{Conclusions and further questions}\label{sec:conclusions}
The finite temperature effective potential plays a fundamental role in understanding the phase structure of quantum field theories, including thermal and quantum corrections with ubiquituous applications in cosmological phase transitions. It was originally developed to describe the  free energy landscape as a function of an order parameter, which is usually a scalar field condensate, by design and construction it is an \emph{equilibrium} concept. However, it is often used in the equation of motion for the order parameter, or ``misaligned'' condensate.

A recent study\cite{veff} of the zero temperature effective potential revealed several important caveats that indicate that using the zero temperature effective potential to describe the dynamics of the condensate is in general unwarranted. Motivated by its importance in cosmology, in this article we focus on understanding if and when the \emph{finite temperature} effective potential is suitable in the \emph{equations of motion} of a homogeneous  condensate.  Extending the  Hamiltonian formulation   we     identify  the finite temperature effective potential with the Helmholtz free energy of the fluctuations around the condensate. This identification has a profound thermodynamic significance: it allows us to  establish a direct relation with the thermodynamic entropy density $\mathcal{S}= -\partial \, V_{eff}[T,\varphi]/\partial T$. Therefore, fundamental thermodynamic properties of the entropy severely restrict  the applicability of the effective potential in a dynamical equation of motion.

When the condensate oscillates around an equilibrium minimum, we find that the entropy is a non-monotonic function of time, whereas if the tree level potential feature symmetry breaking minima, the effective potential and entropy are \emph{complex} when the condensate probes the spinodal region with negative second derivative of the tree level potential. We argue that collisional processes cannot in general maintain local thermodynamic equilibrium unless there is a fine tuning of couplings, and that the time evolution of the condensate leads to a ``freeze-out'' of the density matrix and decoupling from the thermal bath.

A closed quantum system approach based on unitary time evolution yields the correct  and fully renormalized equations of motion for the condensate conserving both energy and entropy, which are manifestly
real and without the caveats of the effective potential. These equations imply an efficient energy transfer mechanism between the condensate and fluctuations as a consequence of profuse \emph{stimulated} particle production via parametric amplification or spinodal instabilities. Particles are produced with non-thermal distribution functions localized in momentum within instability bands either spinodal or parametric, draining energy from the condensate, suggesting the emergence of asymptotic stationary states, the nature of which must be established numerically.

We  focused on obtaining the equations of motion consistently up to one loop, which do not include higher order collisional processes, these  are of paramount importance if re-thermalization is to occur on longer time scales by a redistribution of the created particles. Possible alternative avenues to study these processes would be to implement the effective action approaches introduced in refs.\cite{cornwall,berges}.

Although the study in this article is carried out in Minkowski space time, we expect that many of the lessons will remain relevant in an expanding cosmology. In particular the method to obtain the (causal) equations of motion for the condensate including radiative corrections   may be adapted from those introduced recently\cite{cosmocao} for a different situation within the cosmological context.

\acknowledgements
  S.C. and  D. B. gratefully acknowledge  support from the U.S. National Science Foundation through grants   NSF 2111743 and NSF 2412374.

\appendix

\section{Correlation functions:}\label{app:corre}
In this appendix we summarize the correlation functions of the fluctuating field $\delta(\vx,t)$ that enter in the equation of motion for the mean field $\varphi(t)$. With the quantization of the fluctuation field $\delta(\vx,t)$ given by equation (\ref{delex}), and the initial density matrix  (\ref{rhozeroin}) with the expectation values (\ref{ocuzero}), we find the following correlation functions,

\bea  && \langle \delta^+(\vx,t)\delta^+(\vx',t')\rangle   \equiv  \mathrm{Tr}\, T\Big(\delta(\vx,t)\delta(\vx',t') \Big)\,\rho(t_0)  \nonumber \\ & = &  {\hbar} \, \int \frac{d^3k}{(2\pi)^3}\Bigg\{ \Big[ g^*_k(t) g_k(t')\,e^{-i\vk\cdot(\vx-\vx')}\,n_k(0) + g^*_k(t') g_k(t)\,e^{i\vk\cdot(\vx-\vx')}\Big(1+n_k(0)\Big)\Big]\Theta(t-t') \nonumber \\ & + &  \Big[ g^*_k(t') g_k(t)\,e^{i\vk\cdot(\vx-\vx')}\,n_k(0) + g^*_k(t) g_k(t')\,e^{-i\vk\cdot(\vx-\vx')}\Big(1+n_k(0)\Big)\Big]\Theta(t'-t) \Bigg\} \label{timordef}\eea

\bea  && \langle \delta^-(\vx,t)\delta^-(\vx',t')\rangle   \equiv  \mathrm{Tr}\, \widetilde{T}\Big(\delta(\vx,t)\delta(\vx',t') \Big)\,\rho(t_0)  \nonumber \\ & = &   {\hbar} \, \int \frac{d^3k}{(2\pi)^3}\Bigg\{ \Big[ g^*_k(t) g_k(t')\,e^{-i\vk\cdot(\vx-\vx')}\,n_k(0) + g^*_k(t') g_k(t)\,e^{i\vk\cdot(\vx-\vx')}\Big(1+n_k(0)\Big)\Big]\Theta(t'-t) \nonumber \\ & + &  \Big[ g^*_k(t') g_k(t)\,e^{i\vk\cdot(\vx-\vx')}\,n_k(0) + g^*_k(t) g_k(t')\,e^{-i\vk\cdot(\vx-\vx')}\Big(1+n_k(0)\Big)\Big]\Theta(t-t') \Bigg\} \label{antitimordef}\eea

\bea  && \langle \delta^+(\vx,t)\delta^-(\vx',t')\rangle   \equiv  \mathrm{Tr}\, \Big(\delta(\vx,t)\, \rho(t_0)\,\delta(\vx',t') \Big)\,    =
  \mathrm{Tr}\, \Big(\delta(\vx',t')\,\delta(\vx,t)\, \rho(t_0)   \Big) \nonumber \\
& = & {\hbar} \, \int \frac{d^3k}{(2\pi)^3} \Big[ g^*_k(t') g_k(t)\,e^{i\vk\cdot(\vx-\vx')}\,n_k(0) + g^*_k(t) g_k(t')\,e^{-i\vk\cdot(\vx-\vx')}\Big(1+n_k(0)\Big)\Big] \label{minpludef}\eea

\bea  && \langle \delta^-(\vx,t)\delta^+(\vx',t')\rangle   \equiv  \mathrm{Tr}\, \Big(\delta(\vx',t')\, \rho(t_0)\,\delta(\vx,t) \Big)\,    =
  \mathrm{Tr}\, \Big(\delta(\vx,t)\,\delta(\vx',t')\, \rho(t_0)   \Big) \nonumber \\
& = &  {\hbar} \, \int \frac{d^3k}{(2\pi)^3} \Big[ g^*_k(t) g_k(t')\,e^{-i\vk\cdot(\vx-\vx')}\,n_k(0) + g^*_k(t') g_k(t)\,e^{i\vk\cdot(\vx-\vx')}\Big(1+n_k(0)\Big)\Big] \label{plumindef}\eea

\end{document}